# END-2-END-ARCHITEKTUREN ZUR DATENMONETARISIERUNG IM IIoT
## KONZEPTE & IMPLEMENTIERUNGEN

NOVEMBER 2020

**PREPRINT**

**CHRISTOPH F. STRNADL**

## ZUSAMMENFASSUNG

Das Wertschöpfungspotenzial des Internets der Dinge (*Internet of Things*, IoT), also des Verbindens arbiträrer Objekte mit dem Internet, liegt im Aufschließen und in der nutzenstiftenden Verarbeitung der circa 80 Zettabyte (1 ZB = $10^{21}$ Bytes) an Daten, die von den voraussichtlich 40 Milliarden IoT-Endpunkten produziert werden (Prognose für 2025). Dieser Beitrag behandelt die informationstechnischen Voraussetzungen, die es herzustellen gilt, um dieses Potenzial realisieren zu können. Die Quantität und Heterogenität der besonders im Industrial IoT (IIoT) anzutreffenden Geräte und Maschinen der Industrie bedingen dabei den Einsatz einer typischerweise Cloud-basierten IoT-Plattform zur logischen Bündelung und zum effizienteren Management dieser im Industriekontext unvermeidbaren Komplexität. Stringente nicht-funktionale Anforderungen besonders im Hinblick auf (niedrige) Latenz, (hohe) Bandbreite, Zugang zu großen Rechenkapazitäten sowie Sicherheits- und Vertraulichkeitsaspekte bedingen in der Folge den Einsatz von intermediären IoT-Gateways verschiedenster funktionaler Ausprägungen im Edge-Kontinuum zwischen IoT-Endpunkten und IoT-Plattform in der Cloud. Dies wird an Hand zweier Praxisfälle aus der Perspektive einer Komponentenarchitektur illustriert. Schließlich argumentieren wir, dass dieses klassische Konzept von IoT-Projekten strategisch in Richtung Applikationsintegration (Stichwort: IT/OT-Integration) und API-Management erweitert und über eine Integrations- bzw. API-Management Plattform an die IoT-Plattform angekoppelt werden muss, um im Rahmen eines *End-to-End*-Verständnisses von IoT/IIoT dem innovationsstiftenden Transformationscharakter von Industrie 4.0 gerecht zu werden.

# INHALTSVERZEICHNIS

## ABBILDUNGSVERZEICHNIS

## TABELLENVERZEICHNIS

# 1   EINLEITUNG

Das Konzept des Internets der Dinge, *Internet of Things* (IoT), bezieht sich im Kern auf den Anschluss von arbiträren physischen Objekten an das Internet bzw. an ein geeignetes Kommunikationsnetzwerk [1]. Der Begriff des Objektes wird dabei sehr weit gefasst und beinhaltet alle Arten von Gegenständen von Maschinen oder Geräten (bspw. eine Krananlage oder ein Schweißroboter) über Fahrzeuge (Autos, Schiffe, Eisenbahnzüge und Flugzeuge inklusive Drohnen) bis hin zu allen Ausprägungen von *smart devices,* wie einer *smart watch* oder einem *fitness tracker*, aber auch „intelligenten" Türen oder Kühlschränken im *smart home* oder die Straßenlampen einer *smart city.* Die Auszeichnung der Objekte als „intelligent" (*smart*) hat den technischen Hintergrund, dass sie über hinreichend (elektronische) Rechnerkapazität verfügen müssen (CPU, volatilen und nicht-volatilen Speicher sowie Input/Output-Schnittstellen), um am IoT partizipieren zu können [2].

Über Sensoren[1] greifen diese Objekte dabei Daten über ihren eigenen Zustand oder ihre Umgebung ab, die dann zur weiteren Verarbeitung an andere Knoten im oben angesprochenen Kommunikationsnetzwerk gesendet werden. Insoweit die Objekte auch über sogenannte Aktoren (engl. *actuators*) verfügen, können sie über diese Stelleinrichtungen auch ihre eigene Umgebung beeinflussen[2] [3]. Während typischerweise die Objekte mit entsprechenden IoT-Plattformen (dazu mehr in diesem Beitrag) kommunizieren, ist die oben gegebene Beschreibung der kommunikativen Situation der IoT-Endpunkte völlig allgemein und umfasst u.a. auch die autonome Kommunikation der IoT-Endpunkte untereinander. Man spricht in diesem Falle von Maschine-zu-Maschine-Kommunikation[3] (*machine-to-machine*, M2M) [4][5].

Die Übertragung und Anwendung von Prinzipien, Technologien und Standards des IoT auf die Industrie im Allgemeinen (einschließlich von Transport/Logistik, Gesundheitsindustrie und *smart homes*), ihre Anlagen und Prozesse nennt man dann *Industrial Internet of Things* (IIoT) [7][8]. Dabei werden in der Vollausprägung des Konzeptes[4] sämtliche Anlagen, Maschinen und ihre Steuerungssysteme einschließlich der sich gerade in Produktion befindlichen Werkstücke über den gesamten Produktionsprozess und damit einhergehenden Logistikprozesse hinweg zu einer intelligenten Fabrik[5] (*smart factory*) vernetzt [9]. Die Fokussierung von IIoT auf das produzierende Gewerbe einschließlich seiner produktions- und informationstechnischer Infrastruktur wird dann oft als **Industrie 4.0** bezeichnet[6] [4][10][11]. Diese technische Definition von Industrie 4.0 als „Digitalisierung der Produktion" [10] ergänzt man gerne um eine organisatorisch-strategische Dimension, die dieses Konzept als „grundlegenden Innovations- und Transformationsprozess industrieller Wertschöpfung" mit dem Potenzial, „Organisation und Steuerung der gesamten

---

[1] einschließlich von *tags*, wie bspw. *radio frequency identification* (RFID) *tags*, oft passive Elemente, die erst durch (in der Regel elektromagnetische) Kopplung mit einem weiteren (oft nur Lese-)System in einem begrenzten Ausmaß eine unidirektionale (Tag ⇒ System) bzw. mitunter auch bidirektionale (Tag ⇔ System) Datenkommunikation erlauben.
[2] man denke etwa Stellmotoren, Ventile, Pumpen, aber auch pneumatische Kolben u.v.a.m.
[3] Darüber hinaus war der Begriff M2M um 2010 auch der begriffliche Vorläufer von IoT, ohne dass man damals allerdings der *peer-to-peer* (P2P)-Kommunikation einen so hohen Stellenwert eingeräumt hat [6].
[4] Eine derartige extensive Umsetzung von IIoT ist in der industriellen Praxis derzeit nach Kenntnisstand des Autors deutlich nicht erkennbar.
[5] Man spricht hier auch von einem *cyber-physical system* (CPS)
[6] Diese Unterscheidung zwischen dem allgemeineren Begriff IIoT gegenüber dem eher eng geführten Industrie 4.0 hat sich in der wissenschaftlichen Praxis allerdings noch nicht durchgesetzt. Siehe etwa [8], wo zahlreiche Quellen bspw. IIoT und Industrie 4.0 einfach gleich setzen.

Wertschöpfungskette" auf eine „neue Stufe" zu stellen und damit die vierte industrielle Revolution bewirken zu können, verortet [12].

Die Wertschöpfung durch IoT sowie IIoT beruht dabei auf der konsequenten internen sowie unternehmensübergreifenden nutzenstiftenden Verarbeitung aller verfügbaren Daten (Datenmonetarisierung) [13]. Die Schätzungen für das Jahr 2025 gehen dabei von 11,7 Mrd. IoT-Endpunkten, davon 1,4 Mrd. im Bereich produzierender Industrie und natürliche Rohstoffe [14], bis zu 41,6 Mrd. IoT-Geräten, die jährlich etwa 79,4 ZB (1 Zettabyte = $10^{21}$ Byte = 1 Mrd. TB) produzieren werden [15]. Im Kontext von Industrie 4.0 soll das zu einem Wertschöpfungspotenzial in der Höhe von ca. 3,7 Billionen[7] USD führen [16].

Dieser Beitrag konzentriert sich auf die Darlegung der notwendigen informationstechnischen Voraussetzungen, die jedes Unternehmen schaffen und später aufrecht erhalten muss, um interne und externe Mehrwerte aus den prinzipiellen Konzepten von IIoT bzw. Industrie 4.0 lukrieren zu können. Nach einem architekturorientierten Aufriss des erweiterten *end-to-end* Problemfeldes und der Motivation des für IoT und IIoT gleichermaßen zentralen Konzeptes der IoT-Plattform folgt die Darstellung einem informationslogistischen Ansatz [17]. Die Unterabschnitte gliedern sich daher in Informationsbeschaffung, Informationsbearbeitung und Informationsallokation bzw. -distribution.

Die meisten Konzepte gelten über das „reine" IIoT/Industrie 4.0 hinaus auch für allgemeine IoT-Projekte, ohne dass dies jedoch aus Platzgründen im Einzelfall konkret heraus gearbeitet wird.

## 2 ARCHITEKTUR

### 2.1 Problemstellung

Die primäre technische Aufgabenstellung im IIoT, die für einen Anwendungsfall passenden Geräte, Maschinen und sonstigen IoT-Endpunkte an das Internet anzuschließen, ist zwar einfach zu formulieren, aber alles andere als leicht zu bewerkstelligen. Komplexität trifft man dabei vor allem in den folgenden vier Bereichen an:

- **Komplexität der IIoT-Endpunkte.** Die Heterogenität der potenziell anzuschließenden Maschinen, Geräte, Sensoren, Aktoren oder sonstigen IoT Endpunkte ist ein wesentlicher Komplexitätstreiber auf der Ebene des *shop floors* [4]. Dazu tritt noch die schiere Menge an Geräten, die in typischen Produktionsumgebungen vorhanden sind, wo an größeren Standorten eine Größenordnung von 30.000 – 80.000 Sensoren anzutreffen sind[8].

- **Kommunikationskomplexität.** Während sich im Bereich des (klassischen) Internets mittlerweile ein Kanon an Standards herausgebildet hat (bspw. TCP/IP im Bereich der Netzwerkprotokolle auf Ebenen 3 und 4 und HTTP/REST[9] für Applikationsschnittstellen), ist das im IIoT zur Zeit noch nicht der Fall: Es können bspw. mehr als 120 unterschiedliche Kommunikationsstandards von der physischen Schicht (bspw. die zahlreichen industriellen Feldbus-Protokolle) bis hin zu Applikationsprotokollen (wie etwa CoAP[10] oder MQTT[11]) identifiziert werden [7]. Dazu treten gerade im industriellen Bereich die proprietären Protokolle und Standards der Maschinenhersteller.

---

[7] Hier sind tatsächlich europäische „Billionen" gemeint: 1 Billion = 1.000 Milliarden.
[8] Projekterfahrungen des Autors
[9] *Hypertext transfer protocol / representational state transfer*, das Standardprotokoll in der API-Ökonomie
[10] *constrained application protocol*
[11] *message queuing telemetry transport*

- **Technische Komplexität eines verteilten Systems**. IIoT stellt ein verteiltes System dar, bei dem alle aus dem *distributed computing* bekannten Aspekte, wie bspw. Datenkonsistenz, Durchsatz, Zuverlässigkeit der Knoten bzw. des Netzwerks, Sicherheit u.a.m., insbesondere von den Applikationsentwicklern beachtet werden müssen

- **Anwendungskomplexität**. Produktionsprozesse zeichnen sich gegenüber anderen Unternehmensprozessen durch eine deutlich gesteigerte Komplexität aus, die hier voll auf die Anwendung von IIoT-Prinzipien auf deren Digitalisierung durch schlägt [4][18][19].

## 2.2 Lösungsansätze

Angesichts dieser deutlichen Herausforderung könnte man versucht sein, bei der IT-mäßigen Realisierung bestimmter IIoT-Anwendungsfälle einen vertikalen Lösungsansatz über die Entwicklung bzw. die Beschaffung fachspezifischer (in der Regel Standard-) IoT-Applikationen zu verfolgen. Für den Anwendungsfall «*Condition Monitoring*» wäre das eine «*Condition Monitoring* Applikation», für den Use Case «*Predictive Maintenance*» eine entsprechende «*Predictive Maintenance* Applikation». Dieser Produkt- bzw. Applikations-Ansatz führt jedoch auf Grund der folgenden Faktoren zu einer deutlich ineffizienten Ressourcenallokation:

- **komplexe Integration mit IoT-Endpunkten**: Jede IIoT-Applikation besitzt in aller Regel je individuelle Methoden, sich mit den IoT-Endpunkten zu integrieren, sodass hier trotz zahlreicher IoT-Endpunkte keine Skaleneffekte realisiert werden können. Insoweit die gleichen IoT-Endpunkte an mehrere IIoT-Applikationen angebunden werden müssen, führt dies sogar zu vermeidbarer Mehrarbeit.

- **herstellerabhängige Innovation**: Bei Einsatz von Standardsoftware ist der IoT-Anwender bei Spezifikation und Einsatz von neuen Funktionen sehr stark vom Hersteller der IIoT-Applikation abhängig. Insofern die IIoT-Applikationen anwenderspezifische Adaptionen (*customization*) erlauben, führt das nicht nur zu einem einmaligen Aufwand beim initialen Herstellen der gewünschten (Zusatz-)Funktionalität, sondern auch zu einem laufenden Aufwand bei jedem Versionswechsel.

- **fehlende Agilität**: Ähnlich wie bei klassischen Unternehmensanwendungen führt die zu enge Koppelung zwischen den IIoT-Applikationen und IoT-Endpunkten zu einem Verlust an Flexibilität und Geschwindigkeit [20].

Zur Überwindung dieser Schwierigkeiten greift man auf ein Architekturmuster zurück, das sich schon bei der unternehmensweiten Integration von komplexen IT-Landschaften bewährt hat, nämlich den Einsatz einer **Middleware**[12] [21]. Ein derartiges Software-System ist als „Bindeglied" zwischen

(i) den technischen Komponenten (einschließlich ihrer Software) samt dazwischen vermittelndem Kommunikationssystem und

---

[12] Für diesen Fachbegriff hat sich kein passender deutscher Ausdruck etabliert. Die in [2] ventilierte Bezeichnung als «Infrastruktursoftware» ist deutlich zu weit gefasst und entspricht nicht dem Einsatz in der Praxis, die sich – nach Ansicht des Autors völlig zu Recht – nach wie vor an [21][24] orientiert.

(ii) der „höherwertigen Anwendungslogik", den Applikationen für den Endbenutzer,

konzipiert [22][23]. Es „absorbiert […] die Heterogenität und Komplexität" des Integrationsvorhabens [23][24]. Spezialisiert man dieses Konzept auf IoT, so kann man eine **IoT-Middleware** kurz als „Intermediär zwischen den IoT-Geräten und Applikationen" charakterisieren [25].

Der Einsatz von IoT-Middleware erschließt und ermöglicht damit die folgenden Vorteile bei der Realisierung von IoT-Projekten [22][23]:

- **Standardisierung der Kommunikation** der IoT-Plattform mit den IoT-Endpunkten durch kanonische Übertragungsprotokolle und Datenmodelle[13]
- **Komplexitätsreduktion** durch die Realisierung „höherwertiger Abstraktionen" der IoT-Endpunkte (bspw. durch den „Digitalen Zwilling")
- **Interoperabilität** zwischen IoT-Endpunkten und Applikationen [26]
- **Portierbarkeit** der IoT-Applikationen durch Bereitstellung einheitlicher Programmierschnittstellen (den sogenannten APIs[14])
- **Konzentration auf Anwendungslogik** und den damit verbundenen Mehrwert für die Organisation

Für die software-technische Realisierung dieser IoT-Middleware ergeben sich in der Praxis zwei Möglichkeiten, nämlich die Umsetzung als **IoT-Framework** [27] oder als **IoT-Plattform** [28]. Während ein IoT-Framework als „'halbfertiges' Softwaresystem, bestehend aus einer Vielzahl von auf einander abgestimmten Softwarekomponenten" bezeichnet werden kann, „aus denen mit *relativ geringem* Aufwand [Hervorhebung durch den Autor] ein angepasstes Softwaresystem erstellt werden kann" [23], ist der Kohäsionsgrad bei einer IoT-Plattform um ein Vielfaches höher und der Programmier- und Adaptionsaufwand gegenüber einem IoT-Framework deutlich geringer. Auf diese in der Praxis prävalente Realisierungsform von IoT-Middleware wird ausführlich in Abschn. 3.1 eingegangen.

Die deutlich bessere Wirtschaftlichkeit von IoT-Plattformen gegenüber der unternehmensindividuellen Eigenentwicklung einer IoT-Lösung (etwa auf Basis eines IoT-Frameworks) wird u.a. in [29] bestätigt. Unter Zugrundlegung einer IoT-Lösung mit Anforderungen, die für größere Unternehmen realistisch sind, und einer Betriebsdauer von 5 Jahren für bis zu 5 Mio. IoT-Endpunkte sind die TCO (*total costs of ownership*) einer IoT-Plattform um ca. 40% niedriger als bei einer individuellen Entwicklung.

## 2.3  End-to-end-IoT

In den bisherigen Ausführungen hat sich die räumliche Dimension von IoT/IIoT sehr stark auf die IoT-Endpunkte und ihre Verbindung untereinander sowie mit einer IoT-Plattform konzentriert. Dies entspricht in der Praxis zwar typischerweise dem ersten Schritt („**klassisches IoT**"), vernachlässigt aber die Applikationen und Anwendungen der traditionellen IT wie ERP-, CRM- oder DWH Systeme. Insoweit man diese Systeme ebenfalls (in einem Folgeschritt) in sein IIoT- [sic!] oder Industrie 4.0-Vorhaben miteinbezieht spricht man hier von **IT/OT[15]-Integration** (vgl. Abbildung 1).

---

[13] Das gelingt nur auf Seite der sogenannten *north-bound interfaces.* Die Komplexität auf der *south-bound* Seite der IoT-Middleware kann nicht eliminiert, sondern nur besser sichtbar und damit effizienter steuerbar gemacht werden.
[14] *application programming interface*

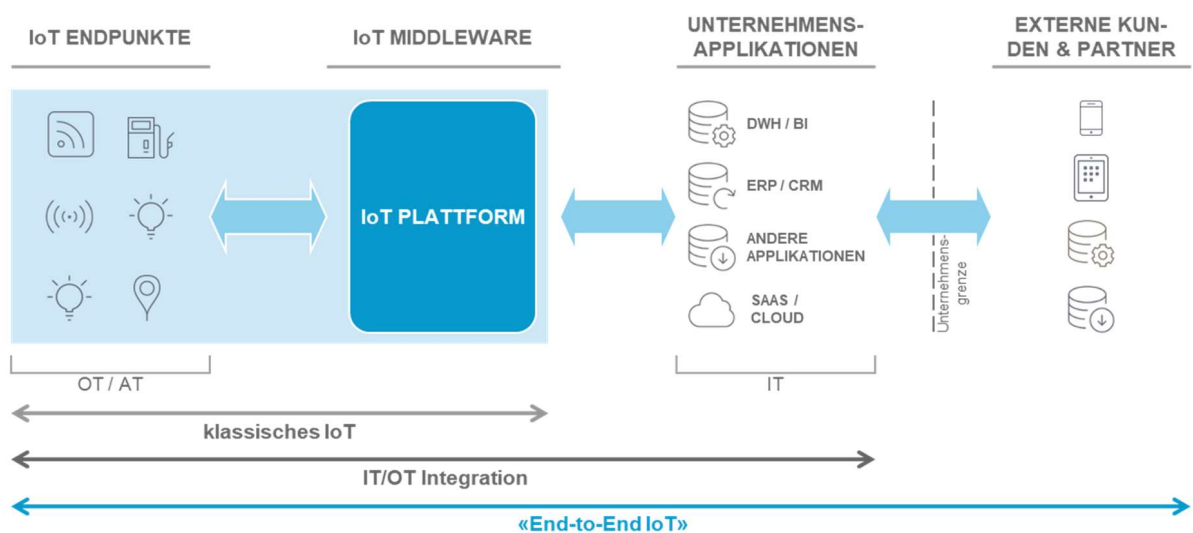

**Abbildung 1.** Dimensionen des IoT: Von klassischer IoT zu *end-to-end*-IoT (eigene Darstellung)

Im Rahmen der API-Ökonomie trifft man aktuell verstärkt die elektronische bzw. informationelle Anbindung unternehmensexterner Organisationen an, die schon in den 1990er Jahren als *electronic data interchange* (EDI) begonnen hat. Integriert man die Teilnehmer dieses sogenannten *extended enterprise* [30] ebenfalls in seine IoT-Strategie, so nennt man diese Konstellation **end-to-end-IoT**.

Die technische Architektur, die die Implementierung einer *end-to-end*-IoT-Lösung erlaubt, wird in den nächsten Abschnitten detaillierter heraus gearbeitet werden.

# 3 INFORMATIONSVERARBEITUNG

## 3.1 IoT-Plattform

### 3.1.1 Definition

Im Gegensatz zu einem bloßen IoT-Framework ist eine IoT-Plattform selbst eine (in der Regel) umfangreiche entwicklungsmäßig abgeschlossene Software-Komponente. Entsprechend ihrem Charakter als *Middleware* zwischen IoT-Endpunkten und IoT-Applikationen bzw. Endbenutzern [31][32] stellt sie einheitliche (IoT-)domänspezifische Kernfunktionen, Services und Schnittstellen zur Verfügung, die von mehreren IoT-Applikationen, die im Kontext der IoT-Plattform ausgeführt werden, geteilt und genutzt werden können [33][34].

---

[15] entsprechend dem englischen Begriff für Automatisierungstechnik (AT), *operational technology* (OT)

Während generische Software-Plattformen dem Entwickler (seltener dem Endbenutzer) Werkzeuge und Funktionen zur Verfügung stellen, die auf den allgemeinen Lebenszyklus von Applikationen abstellen, spezialisieren sich IoT-Plattformen auf die Unterstützung von IoT-Lösungen, die den gesamten Lebenszyklus der IoT-Endpunkte umfassen [35][36]. Da dies insbesondere die oft komplexeren Prozesse von Provisionierung, Management und ggf. Automatisierung von Geräten und Maschinen (*devices*) beinhaltet [36], stellen IoT-Plattformen *in nuce* ein „Ökosystem" zur Verfügung, das die Komplexität, die typischen IoT-Lösungen inhärent ist, von ihren Nutzern, Entwicklern und Endbenutzern, abschirmt [28]. Die damit zur Verfügung gestellte (Software-)Infrastruktur erlaubt dann die „systematische Integration, Aggregation, Speicherung […] und kollaborative Nutzung industrieller Daten" [37]. Die Analyse und Verarbeitung dieser Daten ist dann Basis für die „automatisierten Interaktion zwischen […] Dingen und Prozessen" und ermöglicht in der Folge die Umsetzung „neuer innovativer Geschäftsmodelle wie z.B. Predictive Maintenance" [38].

Technisch lassen sich die obigen Ausführungen wie folgt zu einer exakteren Definition von IoT-Plattform komprimieren[16]:

> **Definition IoT-Plattform.** Unter einer IoT-Plattform versteht man einen kohärenten Satz vorintegrierter Technologien und Werkzeuge, deren Aufgabe es ist, Prozesse und Funktionen im gesamten Lebenszyklus von IoT-Endpunkten und IoT-Applikationen innerhalb eines vordefinierten Rahmens zu vereinfachen, zu standardisieren und zu automatisieren.

Man beachte die zweifache Ausprägung des *life cycle managements* sowohl (i) der angeschlossenen IoT-Endpunkte als auch (ii) der auf der IoT-Plattform ablaufenden IoT-Applikationen. Der Begriff «Technologie» ist in obiger Definition mit Absicht nicht nur auf Software-Technologie beschränkt, da gerade im IoT-Bereich auch die hardware-mäßige Realisation von IoT-Lösungen mit entsprechenden physischen IoT-*Gateways* u.ä. Komponenten notwendig ist. Gleiches gilt für den Begriff «Werkzeuge», wo zwar oft software-technische Werkzeuge (*tools*) angetroffen werden, aber andere Werkzeuge wie bspw. Analyse- oder Vorgehensmodelle im Kontext des informationstechnischen Kerns einer IoT-Plattform hilfreich sind. Der Begriff des «vordefinierten Rahmens» stellt auf die jeder IoT-Plattform inhärente interne Struktur und Architektur ab, mit der sie die (im Endeffekt doch nicht reduzierbare) Komplexität von IoT-Lösungen behandelt.

### 3.1.2   Funktionen und Fähigkeiten

Da sich derzeit weder in der wissenschaftlichen Reflexion [28][36][39][40][41] noch in der unternehmerischen Praxis eine (einzige) universelle IoT-Referenzarchitektur [42] oder eine (einzige) IoT-Plattform- Architektur hat durchsetzen können, sei im Folgenden ein in beiden Domänen gleichermaßen anschlussfähiges Kondensat der **funktionalen Architektur** einer IoT-Plattform vorgestellt.

Während wir die besonderen nicht-funktionalen Anforderungen und die heute erforderlichen Arten der Bereitstellung (*deployment*) von IoT-Plattformen in Abschnitt 3.3 behandeln werden, seien die funktionalen Kernfähigkeiten (*capabilities*) in der nachstehenden Tabelle etwas näher erläutert.

---

[16] ohne den (oft zu beobachtenden) Kategoriefehler zu begehen, Begriffe über ihre Funktionen und Leistungen zu definieren, wie das bspw. bei [40] der Fall ist. Die Frage „Was ist ein Auto" wird eben nicht durch den Satz „Ein Auto hat vier Räder, vier Sitze, einen Motor und ein Lenkrad" beantwortet.

**Tabelle 1.** Typische Funktionen einer IoT-Plattform

| KATEGORIE | BESCHREIBUNG |
|---|---|
| **KERNFUNKTIONEN** | |
| **Device Connectivity** | Herstellen der logischen Verbindung mit den IoT-Endpunkten (*devices*). In zahlreichen Fällen bedeutet dies, dass sich die IoT-Plattform netzwerktechnisch mit einem IoT-Gateway verbindet und dieser dann direkt mit dem IoT-Endpunkt (oft bidirektional) kommuniziert. |
| **Device Management** | Verwaltung und Management des gesamten Lebenszyklus' der angeschlossenen (logischen) IoT-Endpunkte von der Installation, Konfiguration über den laufenden Betrieb bis hin zu Updates (bspw. auch von Gerätesoftware) und der Ausserbetriebstellung. Dies umfasst auch Definition und Verwaltung der im Device integrierten Sensoren, Aktuatoren oder anderer datenproduzierender Teilkomponenten.<br><br>Dazu ist zwingend eine Datenbank der IoT-Endpunkte mit den Metadaten zu Geräten und Maschinen vorzusehen einschließlich typischer Datenbankfunktionen wie Suche, Kategorisierung, Aggregation u.a.m. |
| **Data Management** | Neben der Persistierung der (enormen) Datenmengen, die in der Regel kontinuierlich von den IoT-Endpunkten an die IoT-Plattform übermittelt werden, umfasst dies auch die Definition der entsprechenden Datenmodelle. Da in der Regel IoT-Plattformen nicht als Langzeitspeicher für die von den Geräten gelieferten Daten herangezogen werden, sind entsprechende Mechanismen zum Überführen dieser Daten in andere Speichermedien (bspw. in eine *Data Lake*) mitumfasst.<br><br>Diese *capability* bezieht sich ausdrücklich nur auf statische Daten (*data at rest*). Dynamische Aspekte der Datenverarbeitung bzw. des Datenmanagements finden sich in der Kategorie *Analytics*. |
| **Analytics** | Beginnend mit der visuellen Analyse durch die Benutzer im Rahmen von typischerweise konfigurierbaren Diagrammen und Tabellen alle Arten von Analysewerkzeugen unterschiedlichster Ausprägungen (wie Aggregation, Korrelation) bis hin zur Unterstützung durch spezielle Methoden im Bereich Mustererkennung (bspw. Zeitserienanalyse) oder Algorithmen aus den Domänen künstlicher Intelligenz (KI) und *machine learning* (ML). Diese Fähigkeit umfasst gleichermaßen statische Analysen als auch die Echtzeitanalyse (*real time analytics*) von Datenströmen (Abschn. 5.1). |
| **Events & Rules** | Definition von bestimmten Ereignissen mit Hilfe geeigneter (Geschäfts-)Regeln (bspw. Temperatur im kritischen Bereich) auf Basis |

| | der von den IoT-Endpunkten übermittelten Daten sowie das Einrichten und die automatische Detektion derselben. |
|---|---|
| **IoT Application Support** | Erlaubt und unterstützt die Entwicklung, den Einsatz und den Betrieb von eigener, plattformunabhängiger Fach- und Anwendungslogik. Das entsprechende Kontinuum von einfacher Konfiguration der IoT-Plattform bis hin zur Orchestrierung von Microservices wird in Abschn. 3.1.3 näher erläutert. |
| **ERWEITERTE FUNKTIONEN** ||
| **Integration** | Bidirektionale Integration der IoT-Plattform mit den übrigen unternehmensinternen IT-Systemen und Applikationen sowie (besonders für *end-to-end*-IoT/IIoT) externen Organisationen des *extended enterprise* (Abschn. 5.2). Dies erfolgt in vielen Fälle auf Grund der Komplexität nicht mehr direkt in der IoT-Plattform selbst sondern über deren Kopplung mit entsprechenden Integrations- oder API-Plattformen. |
| **Process Automation** | Die Automatisierung von Prozessen mit und ohne humanen Akteuren in Form der *ex ante* Orchestrierung der den Prozess konstituierenden Aktivitäten. Dies kann von der Automatisierung von Aufgaben an einer einzelnen Maschine (*task automation*) bis hin zu maschinen- und applikationsübergreifenden Prozessen mit zahlreichen Rollen reichen. Je komplexer die zu automatisierenden Prozesse sind, umso eher wird diese Aufgabe jedoch in einem separaten, an die IoT-Plattform gekoppeltes *Business Process Management System* (BPMS) realisiert. |
| **IoT Solution Accelerators** | Im Kern sind IoT-Plattformen in der Regel *qua* Konstruktion als (horizontale) technische Plattform branchen- und industriezweigagnostisch. Manche IoT-Plattformen unterstützen dennoch Konfiguration und Bereitstellung spezifischer vertikaler (branchenabhängigen) Lösungen (*solutions*) oder Rahmenwerken derartige Lösungen (*solution accelerators*). |
| **Administration** | Als selbständige (IT-)Anwendung benötigt die IoT-Plattform geeignete Funktionen zu ihrer Administration etwa hinsichtlich der Benutzer und ihrer Rechte, der technischen Konfiguration (Lastmanagement, Sicherheitsmanagement u.v.a.m.) oder der unterschiedlichen Mandanten (Abschn. 3.3.2) insofern die IoT-Plattform überhaupt mandantenfähig ist. |

### 3.1.3   Unterstützung von IoT-Applikationen

Entsprechend der Definition einer Software-Plattform im Allgemeinen und einer IoT-Plattform im Speziellen (Abschn. 3.1.1) erwartet der Benutzer derselben die entsprechende Funktionalität, eigene Anwendungs- bzw. Fachlogik entwickeln, ausführen und managen zu können. In der Praxis hat sich dabei ein Spektrum von typischen Fähigkeiten heraus gebildet, das sich von der bloßen Konfiguration über Anpassungen (*customization*) bis hin zur software-technischen Entwicklung von Anwendungen erstreckt.

Nach dem Definieren und Verbinden der IoT-Endgeräte mit der IoT-Plattform steht das Darstellen der von den Geräten empfangenen Daten in geeigneten Visualisierungsformen (*dashboards*) sowie das Erstellen von (einfachen) Berichten und fachlichen Regeln (*smart rules*), die dynamisch auf Ereignisse bzw. Daten der IoT-Endpunkte reagieren, im Vordergrund. (*Power*) Usern erlauben die typischen IoT-Plattformen das in Form von bloßer **Konfiguration**, also ohne die Notwendigkeit, selbst Code erstellen zu müssen (vgl. den hellen/orangen Rahmen in Abbildung 2 ganz links).

Wenn sich die Möglichkeiten der Änderung der IoT-Plattform durch Konfiguration erschöpfen, erlauben zahlreiche IoT-Plattformen **Anpassungen** und Erweiterungen durch das Erstellen bzw. den Einsatz von Code-Komponenten geringerer Komplexität, wie bspw. *Plug-Ins* für die graphische Darstellung oder die Entwicklung eigener Geschäftsregeln auf Basis entsprechender *Rules Languages* (vgl. die hellen/orangen Elemente in Abbildung 2). Dies kann sich soweit erstrecken, dass die Entwicklung eigener **plattform-spezifischer IoT-Applikationen** unterstützt wird. In aller Regel setzt diese Option Kenntnisse und Wissensstand eines IoT-Plattform-Entwicklers voraus, der die Details des je individuellen und plattform-spezifischen Applikationssupports-APIs kennt.

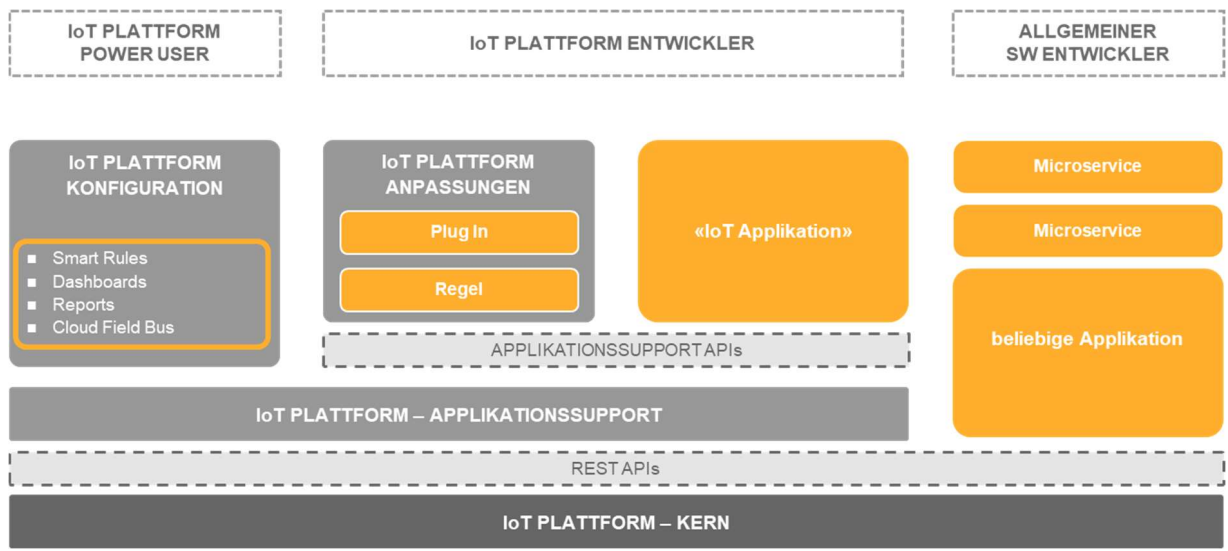

**Abbildung 2.** Dimensionen der Unterstützung von IoT-Applikationen (eigene Darstellung).

Zuletzt ermöglicht die Mehrzahl der IoT-Plattformen in unterschiedlicher Ausprägung, dass beliebige **plattform-externe Software-Komponenten** (Applikationen oder Microservices) über entsprechende (in der Regel: REST) Schnittstellen auf ihre Kernfunktionen zugreifen können. Wesentliche Unterschiede bestehen dahingehend, ob diese Komponenten noch im Kontext der Plattform selbst (also bspw. durch Plattformfunktionen) aktiviert oder bereit gestellt werden oder komplett von ihr getrennt ablaufen.

## 3.2 Informationsverarbeitung im Edge-Bereich

### 3.2.1 Treiber

In zahlreichen IoT-Anwendungsfällen ist die direkte Anbindung der IoT-Endgeräte an eine IoT-Plattform in der Cloud in Form eines PaaS- oder SaaS-Konzeptes auf Grund der Einfachheit und der technischen Vorteile, wie bspw. Skalierbarkeit, Hochverfügbarkeit und effizientem (weil zentralem) Management die Architektur der Wahl. Für etliche IoT-Applikationen jedoch sind die damit verbundenen Nachteile in den folgenden **nicht-funktionalen Anforderungen** so groß, dass eine reine Cloud-Architektur nicht eingesetzt werden kann [43][44][45][46][47][48][49]:

- **niedrige Latenzzeiten:** Die zeitnahe Steuerung von Maschinenanlagen oder sich bewegenden Geräten setzt voraus, dass die Zeit zwischen dem Absenden der Daten an die IoT-Plattform bis zum Empfangen einer entsprechenden Reaktion gering bleibt. Dies kann mit einer IoT-Plattform in der Cloud auf Grund der (zu langen) Übertragungszeiten im Netzwerk nicht realisiert werden.
- **hohe Bandbreite:** Zahlreiche IoT-Endgeräte produzieren so viele Daten (bspw. Videoüberwachung, komplexe Industrieanlagen, Positionsdaten mobiler Endgeräte), dass eine zeitnahe Übermittlung aller dieser Daten an eine IoT-Plattform in der Cloud wirtschaftlich und technisch nicht vertretbar ist.
- **zu geringe Rechenkapazität im IoT-Endpunkt**: Nicht alle IoT-Endgeräte verfügen über die notwendigen Rechenkapazitäten (CPU, Speicher, aber auch Energieversorgung bzw. Batteriekapazitäten), um die durch die ersten beiden Anforderungen notwendig gewordenen Datenverarbeitungsschritte selbst durchführen zu können.
- **Sicherheit & Vertraulichkeit**: Liegt die IoT-Plattform in der Cloud, ist eine durchgängige bzw. genügend durchlässige Netzwerkverbindung zu derselben Voraussetzung. Dies ist besonders im IIoT-Umfeld — oft aus dem Blickwinkel von (fehlender) Sicherheit bzw. Vertraulichkeit — nicht immer gegeben. Dazu tritt ggf. noch mangelndes Vertrauen in Hinblick auf Sicherheit und Vertraulichkeit in die zentrale IoT-Plattform in der Cloud selbst [44][48][49][50].
- **rechtliche Anforderungen**. Mitunter führen auch rechtliche Vorschriften und Gesetze dazu, dass der Einsatz eine reiner Cloud-Lösung nicht möglich ist: *off-shore* Windturbinen müssen bspw. mit lokaler Steuerungslogik (d.h. direkt am Mast) ausgestattet sein; Patientendaten von hospitalisierten Personen müssen mitunter lokal direkt im Spital (also *on premise*) gespeichert werden.

Die technische Lösung dieser mit einem ausschließlich Cloud-basierten Ansatz konfligierenden Anforderungen besteht darin, entsprechende aktive Informationsverarbeitungsressourcen (Rechenkapazitäten, Speicher, Kommunikationsmodule, Anwendungen bzw. Services und Managementfunktionen) näher an die IoT-Endgeräte bzw. IoT-Benutzer zu bringen, und wird „**Edge Computing**"[17] genannt [46][48]. Der Begriff „Edge" bezieht sich dabei auf den Rand des Netzwerkes, das in die Cloud (i.d.R. zur dort situierten IoT-Plattform) führt [43][44][51].

---

[17] Eine deutschsprachige Übersetzung für die Verwendung von "Edge" im Kontext von IoT hat sich (noch) nicht herausgebildet.

Dies wird öfter mit identischer Bedeutung auch als „**Fog Computing**" bezeichnet [43][52][53][54], auch wenn — mit den besseren Gründen in der Sicht des Autors — diese Bezeichnung für diejenige Unterkategorie von Edge Computing reserviert bleiben sollte, die einer sogenannten *Cloud-nativen Architektur* folgt (bspw. durch entsprechende Virtualisierung, Hierarchisierung oder sonstige Distribution) [44][55][56][18]. Der Fokus auf die der klassischen Cloud nachgebildete Art und Weise der Bereitstellung von Edge Computing-Infrastruktur [43] verleitet manche sogar, die Bezeichnung „**Edge Cloud**" zu verwenden [44].

### 3.2.2   Das Edge-Kontinuum

Als „Brücke zur Realisierung der Konvergenz zwischen dem physischen Raum und dem virtuellen Raum im IoT-Paradigma" [48] umfasst Edge-Computing den gesamten „Pfad zwischen Datenquellen und Cloud-Rechenzentren" [43]. Es wäre daher unzweckmäßig, Edge-Computing an einer einzigen physischen Lokation zu verorten; man sollte sich die „Edge" eher als **Edge-Kontinuum** mit zahlreichen intermediären Positionen zwischen dem IoT-Endgerät auf der einen (in Abbildung 3 auf der linken) Seite des Spektrums und der Cloud (typischerweise eines Hyperscalers) auf der gegenüber liegenden (rechten) Seite des Spektrums vorstellen [50][57]. Insoweit man die Cloud in dieses Kontinuum miteinbezieht, kann man vom **Cloud-to-Edge-Kontinuum** (**C2E**) sprechen [58][59][60][61].

Man beachte in der Abbildung 3, dass die Bezeichnungen der einzelnen Positionen innerhalb des Edge Kontinuums keineswegs standardisiert sind und als primäres Ordnungskriterium eigentlich nur die (physische) Entfernung vom IoT-Endgerät einigermaßen aussagekräftig ist.

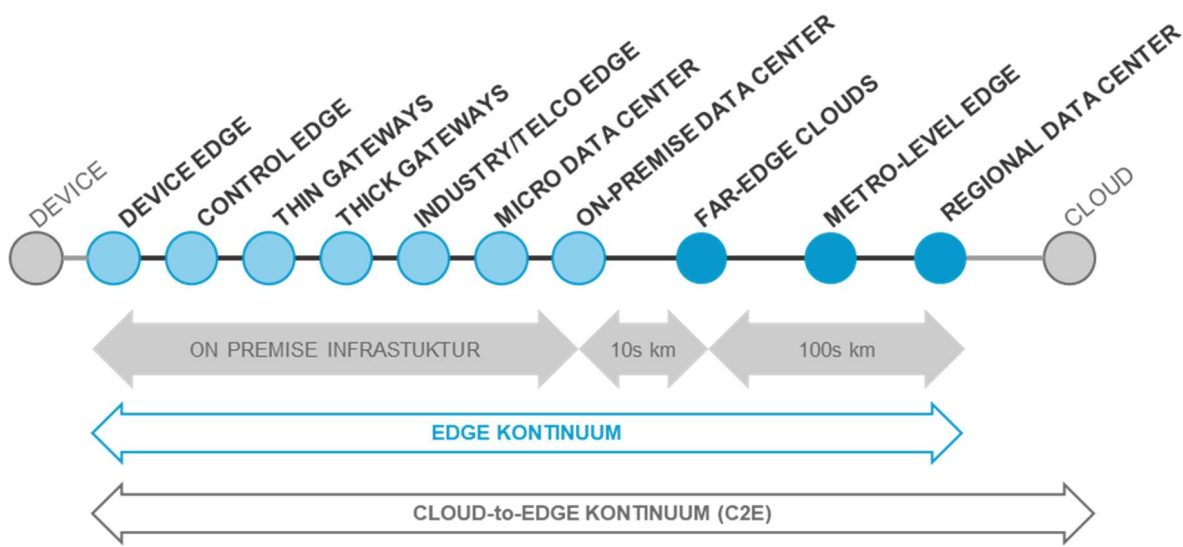

**Abbildung 3.** Edge-Kontinuum (eigene Darstellung)

Dem Edge-Kontinuum auf Ebene der (Edge-)Infrastruktur entspricht analog ein Kontinuum, wie die nicht-funktionalen Anforderungen (Abschn. 3.2.1), die die Architektur überhaupt erst in

---

[18] In diesem Sinne wird Fog Computing zutreffenderweise als "*cloud computing closer to the ground*" bezeichnet [55].

Richtung Edge zwingen, typischerweise realisiert werden können. Dies ergibt sich aus der betriebswirtschaftlichen Forderung nach Effizienz von Ressourceneinsatz und Aufwand, diese zu betreiben und zu steuern. Da die Kardinalität der Instanzen von der Cloud in Richtung IoT-Endpunkt (*device*) massiv ansteigt, wird man nur diejenigen Funktionen in die Edge verlagern, die für die konkrete Realisierung der Anforderungen eines IoT-Projektes notwendig sind. Die (im Endeffekt dann verteilte) IoT-Plattform muss daher die Bereitstellung entsprechender Fähigkeiten (*capabilities*) auf unterschiedlich mächtigen Umgebungen, die in unterschiedlichen Mengen auftreten, erlauben. Abbildung 4 zeigt dies exemplarisch für ein Edge-Kontinuum mit vier Positionen zwischen IoT-Gerät und der Cloud.

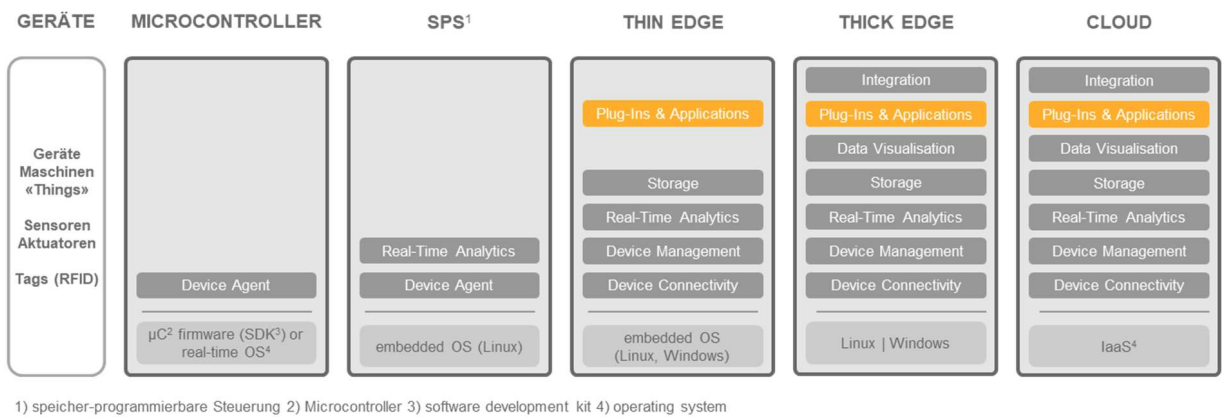

**Abbildung 4.** IoT-Plattform-Funktionen im Edge-Kontinuum (eigene Darstellung)

Die orangen (hellen) Elemente in der Abbildung stehen dabei für die Funktionalität, individuelle Software-Komponenten von einfachen „Plug Ins" (etwa für das User-Interface oder für Geschäftsregeln von geringer Komplexität) bis hin zu „Applikationen" (einschließlich von Microservices) auf diesem Edge-Device bereit stellen zu können.

### 3.2.3   Anwendungsfall 1 – IoT-Gateways

Ein prävalenter Anwendungsfall von Edge Computing ist der Einsatz von sogenannten «**IoT-Gateways**», die zwischen IoT-Endgerät und IoT-Plattform (in der Cloud) geschaltet werden, um (i) (Kommunikations-)Protokolle zwischen den beiden zu übersetzen (Abschn. 2.1) und (ii) eine sicherheitstechnische Abschottung zwischen IoT-Endpunkt und dem Internet zu erreichen. Physisch befinden sich diese IoT-Gateways oft in allernächster Nähe zu den IoT-Endgeräten, also direkt vor Ort an der Produktionsstätte.

Im nachstehenden Praxisbeispiel (vgl. Abbildung 5) werden Industriekompressoren, die in industriellen Produktionsanlagen für komprimierte Luft unterschiedlicher Drucke und Güteklassen[19] sorgen, über IoT-Gateways an eine zentrale IoT-Plattform angeschlossen. Die Kompressoren unterschiedlicher Typen setzen dabei zahlreiche Industrieprotokolle — sogenannte Feldbusse wie Modbus, CAN[20] bus und zahlreiche andere — zur Kommunikation mit dem IoT-Gateways ein. Die

---

[19] bspw. mit Einsatz von Ölfiltern oder ohne
[20] *controller area network*

IoT-Gateways wiederum kommunizieren über Mobilfunk (GSM, UMTS, 5G) mit der IoT-Plattform und nutzen deren Standardschnittstellen, wie bspw. HTTP/REST oder MQTT[21] (vgl. „1. Generation"). Dies setzt naturgemäß voraus, dass sie über entsprechende Software-Komponenten für diese Protokolle verfügen, wie etwa einen MQTT-Client.

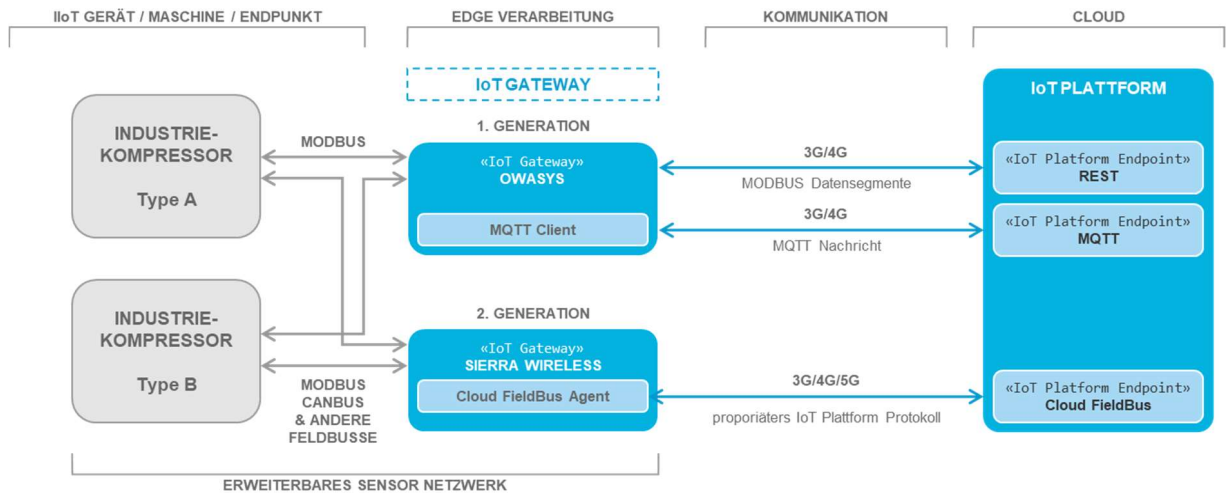

**Abbildung 5.** Anwendungsfall 1 – IoT-Gateways (eigene Darstellung)

Während die Protokollübersetzung von IoT-Gateways der ersten Generation nur lokal konfiguriert werden konnte, sind IoT-Gateways der zweiten Generation „*cloud native*": Hier kann das IoT- Gateway und insbesondere seine Übersetzungsregeln über die IoT-Plattform in der Cloud zentral konfiguriert und verwaltet werden. Dieses Konzept wird hier Cloud Fieldbus genannt[22]. Auch dies wird über eine entsprechende lokale, d.h. am IoT-Gateway selbst laufenden Software-Komponente unterstützt, den «Cloud Fieldbus Agent». Dieser Architekturansatz hat beim Nachteil der Plattformspezifität den Vorteil, dass dadurch das *Cloud-to-Edge* (C2E)-Kontinuum noch besser unterstützt wird als dies über offene Standardprotokolle der Fall ist[23]. Diese Abwägung (*trade off*) zwischen Standardisierung *versus* Plattformabhängigkeit zu Gunsten von Effizienzgewinnen ist jeweils im Einzelfall zu entscheiden.

### 3.2.4  Anwendungsfall 2 – Edge-Analytics

Ein geradezu paradigmatischer Anwendungsfall von Edge-Computing besteht im Durchführen von analytisch-orientierter Informationsverarbeitung nahe den IoT-Endpunkten, dem sogenannten «**Edge-Analytics**». Primärer Treiber dafür, dass überhaupt Edge-Computing eingesetzt und nicht alles einfach in der Cloud berechnet wird, ist die notwendige niedrige Latenzzeit, die es zu erreichen gilt. Als sekundärer Nutzeneffekt sei die durch diese frühzeitige Aggregierung erzielbare

---

[21] *message queuing telemetry transport*, ein nachrichten-orientiertes Client/Server Protokoll
[22] Da es sich hier um kanonisches Protokoll für typische industrielle Feldbusse, wie bspw. Modbus, Profibus, Controller Area Network (CAN), handelt, das für die Kommunikation zwischen Edge-Device (hier einem IoT-Gateway) und der IoT-Plattform in der Cloud optimiert ist.
[23] Vorteile ergeben sich in funktionaler (vor allem durch das Ausdehnen plattformspezifischer Abstraktionen auf die Edge-Devices) und nicht-funktionaler Hinsicht, wie bspw. Bandbreiten- und Latenzreduktion, Sicherheit, Zentralisierung der Verantwortlichkeiten und Erhöhung der Agilität.

Entlastung des Netzwerks (Bandbreite und Übertragungsmenge) nicht unerwähnt (Abschn. 3.2.2).

Das nachstehende Architekturbeispiel beschreibt eine typische Installation einer Lackierstation in einem (bspw. Automobil-)Werk mit jeweils ca. 6-12 Lackierrobotern. Der Anwendungsfall bezieht sich auf die kontinuierliche Überwachung aller Roboter einer Station, um rechtzeitig drohende Fehlfunktionen am Roboter (bspw. eine Verstopfung oder zu geringer Luftdruck an einer Düse oder Fehler in der Farbversorgung) erkennen und korrektive Maßnahmen einleiten zu können, um Schäden am Roboter selbst oder am Werkstück zu vermeiden. In IoT-Plattformen wird dies über die Funktion «Streaming Analytics» mit entsprechenden (individuellen) Geschäftsregeln zu diesem Anwendungsfall erreicht.

Derartige Lackierroboter weisen einen typischen Befehlszyklus (*duty cycle*) in der Größenordnung von 4-20 ms auf. Reaktionen auf eine (drohende) Fehlfunktion müssen daher in einem entsprechend kurzen Zeitintervall (hier < 500 ms) zurück gespiegelt werden, will man Schäden verhindern. Da eine derart niedrige Latenz technisch nicht zu erreichen ist, wenn man die Analyse in einer in der Cloud situierten IoT-Plattform vornimmt, wird in diesem Fall eine «**IoT Edge- Plattform**» mit der entsprechenden «Streaming Analytics»-Funktion eingesetzt, die physisch direkt an der Produktionsstätte betrieben wird (siehe Abbildung 6).

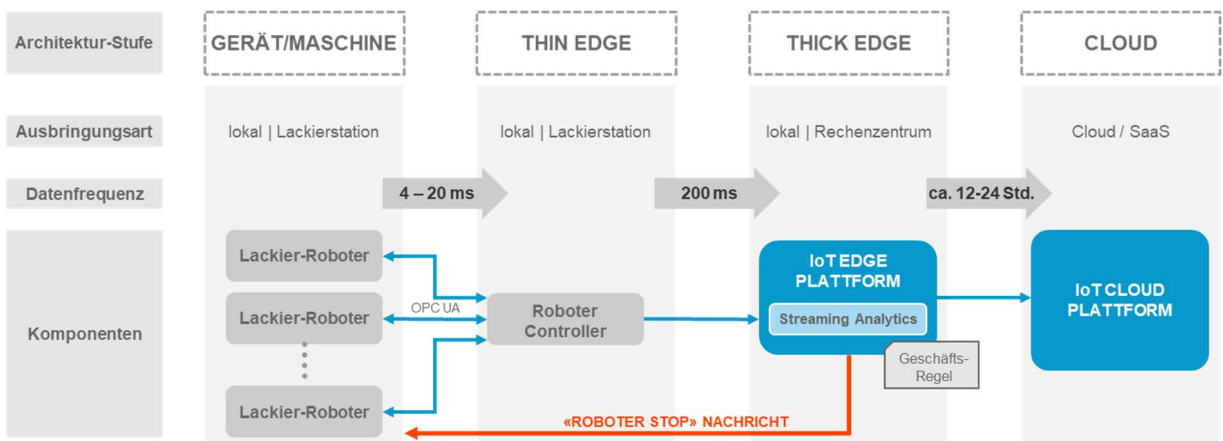

**Abbildung 6.** Anwendungsbeispiel 2 – Edge-Computing

Der «Roboter Controller», der die Lackierroboter direkt ansteuert und überwacht, fasst die Echtzeitdaten der gesamten Lackierstation zusammen und leitet sie alle 200 ms an die IoT-Edge-Plattform weiter. Stellt die «Streaming Analytics»-Geschäftsregel auf Basis der laufenden Analyse des kontinuierlichen Datenstroms fest, dass mit einer gewissen Wahrscheinlichkeit eine Fehlfunktion bei einem bestimmten Lackierroboter zu erwarten ist, sendet sie einen entsprechenden korrektiven Befehl direkt an den betroffenen Lackierroboter. Im angeführten Anwendungsfall hat sich in der Praxis gezeigt, dass bereits ein einfacher STOP-Befehl an den Roboter in der Lage ist, in der Mehrzahl der Fälle Fehler zu verhindern. Die eigentliche Fehlerbehebung wird hier noch manuell durch entsprechende Spezialisten vor Ort durchgeführt.

Aggregierte Daten werden dann von der IoT Edge-Plattform in regelmäßigen Intervallen an die IoT Cloud-Plattform übergeleitet, um auch zentral einen Überblick über die gesamte IoT-Lösung zu bekommen.

Weitere Treiber in Richtung Edge-Computing sind in diesem Kontext neben der Latenz auch die höhere **Sicherheit** des lokalen Edge-Devices gegenüber Hacker-Attacken bzw. Schadsoftware sowie die **Vertraulichkeit** der detaillierten Maschinendaten der Lackierroboter. Etliche Organisationen wolle diese „intimen" Daten ihrer Geräte nicht bzw. höchstens in einer aggregierten Form an die Cloud-Instanz einer IoT-Plattform übergeben.

## 3.3 Nicht-funktionale Anforderungen

Von den zahlreichen nicht-funktionalen Anforderungen an eine IoT-Plattform seien in diesem Abschnitt drei Qualitätsattribute besonders behandelt, die im Kontext von IoT-Projekten eine höhere Wichtigkeit besitzen als dies bei anderen („klassischen") IT-Vorhaben der Fall sein mag.

### 3.3.1 Skalierbarkeit

Typische industrielle Produktionsanlagen umfassen in der Größenordnung von $10^4$-$10^5$ Sensoren, $10^3$ Steuergeräte und -Konsolen und (bei voller Digitalisierung) etwa $10^2$ IoT-Gateways, die täglich 100-300 GB an Daten produzieren (Abschn. 5.1). Angesichts dieser Volumina versteht es sich von selbst, dass die initialen IoT-Projekte mit einer deutlich geringeren Anzahl an IoT-Endpunkten beginnen, die oft in der Größenordnung von $10^1$ - $10^2$ und entsprechend niedrigeren Datenmengen liegen.

Um hier in späteren Umsetzungsphasen der IoT-Strategie nicht die erstmalig gewählte IoT-Plattform wechseln zu müssen, ist das rigorose Testen der Performanz bei einer deutlich höheren Auslastung zwingend erforderlich, um die später gewünschte (und von Herstellern naturgemäß gerne zugesagte) hohe **Skalierbarkeit** zu gewährleisten. In der Praxis bedeutet dies einen nicht unerheblichen Aufwand, der sich aber in späteren Phasen der Umsetzung der IoT-Strategie in einer deutlich erhöhten Leistung und nachhaltig besseren *User Experience* niederschlägt.

### 3.3.2 Mandantenfähigkeit

Unter **Mandantenfähigkeit** versteht man, dass „in einem System für mehrere rechtlich getrennte Unternehmen (= Mandanten[24]) Daten gespeichert und verarbeitet werden, ohne dass die Unternehmen ihre Daten einander preisgeben müssen. […] [Es] werden die Zugriffsrechte und Sichtbarkeit der Daten geregelt, wobei es auch mandantenübergreifende, d.h. für alle [oder für eine bestimmte Untergruppe, Anm. Strnadl] sichtbare Daten geben kann." [62]. Das erlaubt es jedem Betreiber einer IoT-Plattform, die mit dem Betrieb einer einzigen (physischen) Instanz verbundenen Kostenvorteile zu lukrieren, ohne die Vertraulichkeit und Integrität der Daten der Mandanten kompromittieren zu müssen.

Dies ist hinreichend und seit langem bspw. im Bereich der ERP-Systeme bekannt, wo auf einer physischen Installation von SAP R/3 mehrere unterschiedliche Mandanten implementiert werden können; eine Fähigkeit, die für jeden Konzern bzw. jede Unternehmensgruppe essentiell ist, die aus naheliegenden Gründen nicht für jedes Tochterunternehmen eine eigene Instanz betreiben und managen möchten.

---

[24] im Englischen *tenant*

Da Unternehmen IoT-Plattformen gerne in Form von SaaS oder PaaS beziehen, die von einem Dritten betrieben werden, sei an dieser Stelle eine spezielle Ausprägung der Mandantenfähigkeit — **hierarchische Mandantenfähigkeit** — erwähnt. Dabei stellt der Betreiber der IoT-Plattform seinen Kunden (= Mandanten) nicht nur eine einfache Einzelmandanten-Instanz der IoT-Plattform zur Verfügung, sondern ggf. auch eine selbst wieder multimandantenfähige Instanz. Damit werden die Kunden in die Lage versetzt, ihrerseits ihren eigenen Tochterunternehmen oder anderen Dritten (vor allem ihren Kunden) Mandanten innerhalb ihrer Multimandanteninstanz zu allozieren. Die Mandantenfähigkeit einer Instanz kann dadurch hierarchisch vererbt und weitergegeben werden (sofern gewollt).

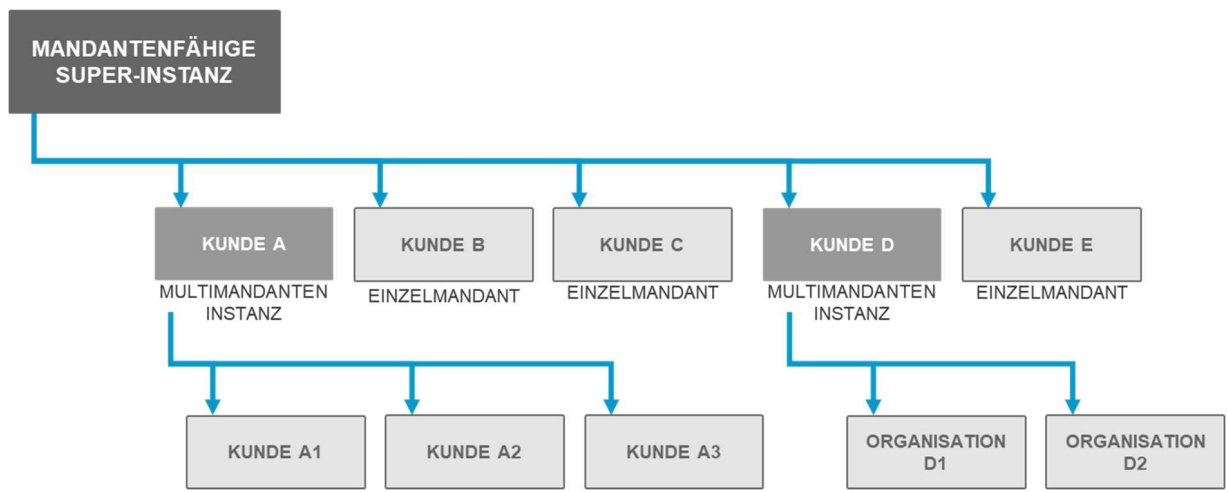

**Abbildung 7.** Hierarchische Mandantenfähigkeit einer IoT-Plattform (eigene Darstellung)

In Abbildung 7 betrifft das die Kunden A und D, die mit Hilfe ihrer Multimandanteninstanz der IoT-Plattform ihren Kunden (im Falle von «A») oder anderen Organisationen (im Falle von «D») je individuelle Einzelmandanten zuordnen können.

Mit Hilfe dieser Eigenschaft wird u.a. der Betrieb von großen B2B[25] IIoT-Plattformen wie die ADAMOS-Plattform [63] möglich gemacht. Die Multimandantenfähigkeit der dabei zum Einsatz gebrachten IoT-Plattform geht dabei so weit, dass nicht nur die Sichtbarkeit bzw. Zugänglichkeit von Daten pro Mandant je individuell definiert werden kann, sondern auch Applikationen bzw. Microservices, die im Kontext der IoT-Plattform laufen und dort bereitgestellt werden, je Mandant individuell freigeschaltet und genutzt werden können (oder eben nicht).

### 3.3.3   Bereitstellungsoptionen (*deployment*)

IoT-Plattformen werden in der Regel in Form von PaaS (*platform as a service*) oder SaaS (*software as a service*) angeboten und genutzt. Welche Hardware-Plattform (durchaus im Sinne von IaaS, *infrastructure as a service*) vom Anbieter oder Betreiber der IoT-Plattform letztlich genutzt wird, bleibt dabei für Kunden wie Endanwender gleichermaßen in vielen Fällen beabsichtigterweise verborgen und nicht beeinflussbar.

---

[25] *business-to-business*

Dies ist aber nicht in allen Fällen wirtschaftlich zweckmäßig oder rechtlich möglich:

Man denke bspw. an länderspezifische gesetzliche Vorgaben betreffend die Datenspeicherung (besonders von personenbezogenen Daten) auf Servern, die zwingendermaßen im eigenen Hoheitsgebiet zu situieren sind (bspw. Russland oder China). In solchen Fällen kann der IoT-Plattform-Betreiber nicht mehr die IaaS-Dienste der westlichen Hyperscaler (Amazon, Microsoft, Google) als Infrastruktur für seine Plattform nutzen, sondern muss etwa auf Alibaba Cloud oder Tencent Cloud als Hyperscaler ausweichen.

In anderen Fällen lassen die (u.U. Multi-)Cloud-Umgebungen der Nutzer der IoT-Plattform es zweckmäßiger erscheinen, einen ganz bestimmten Hyperscaler als IaaS-Dienstleister heran zu ziehen, bspw. wenn es um (zu vermeidende) Latenzen oder Kostensynergien geht. In eine ähnliche Kategorie fallen Installationen der IoT-Plattform (oder ihrer Teilkomponenten) *on premise* (auf lokale Infrastruktur) oder in bestimmten Lokationen im Edge-Kontinuum (Abschn. 3.2.2), die auf Grund der nichtfunktionalen Anforderungen zwingend notwendig werden können.

Schließlich gibt es Situationen (besonders im Telekommunikations- bzw. CSP[26]-Segment), bei denen ein Hersteller einer IoT-Plattform diese an einen Dritten zum selbstständigen Betrieb unterlizensiert und dieser Betreiber bestimmte Präferenzen für seinen IaaS Dienstleister hat.

Allen Konstellationen gemeinsam ist die Anforderung an die IoT-Plattform, möglichst unabhängig von der zu Grunde liegenden (Hardware- oder Virtualisierungs-)Infrastruktur bzw. den IaaS-Diensten zu sein. IoT-Plattformen unterstützen daher *idealiter* das Konzept *one platform – deploy anywhere*.

# 4      INFORMATIONSBESCHAFFUNG

## 4.1   IoT-Gateways und andere Architekturmuster

Die grundlegende Aufgabe der Datenbeschaffung beim IoT besteht darin, Geräte, Maschinen, Dinge (*Things*) und andere IoT-Endpunkte über ein geeignetes Kommunikationsnetzwerk an eine IoT-Plattform anzubinden. Gewöhnlich wird die Kommunikation zunächst in der Orientierung Gerät ⇨ IoT Plattform verlaufen; ab einem gewissen IoT-Reifegrad der Organisation wird dies typischerweise bald in eine bidirektionale Kommunikation Gerät ⇔ IoT-Plattform erweitert werden[27]. Diese Aufgabe ist zwar leicht formuliert, in der Praxis aber nicht nur nicht einfach, sondern auf Grund der Komplexität (vgl. die Ausführungen weiter oben) ein deutlicher Hemmschuh für die Proliferation von IoT-Anwendungsfällen. Dabei kann man im Wesentlichen zwei Strategien unterscheiden, den Green Field und den Brown Field Ansatz.

Beim **Green Field-Ansatz** geht es um die Anbindung von neu anzuschaffenden Geräten, die man idealerweise so auswählt (bzw. auswählen kann), dass sie nativ die von der IoT-Plattform akzeptierten Protokolle unterstützen. Im industriellen Umfeld ist dies vor allem OPC UA[28]; sonst wird man HTTP/REST oder MQTT bevorzugen, bzw. bei der Anbindung von Geräten, die unter knapper Energieversorgung operieren, LWM2M[29].

---

[26] *communication services provider*
[27] z.B.: IoT-Plattform sendet Befehle an das Gerät bis hin zu einem Update der Software auf dem Gerät
[28] *Open Platform Communications – universal architecture.* Ein service-orientiertes Protokoll zur Anbindung und Steuerung von Maschinen.

Für existierende Geräte oder solche Neugeräte, die nicht standardmäßig (*out of the box*) an die IoT-Plattform angebunden werden können – der sogenannte **Brown Field-Ansatz** – haben sich zahlreiche Architekturmuster herausgebildet (vgl. Abbildung 8).

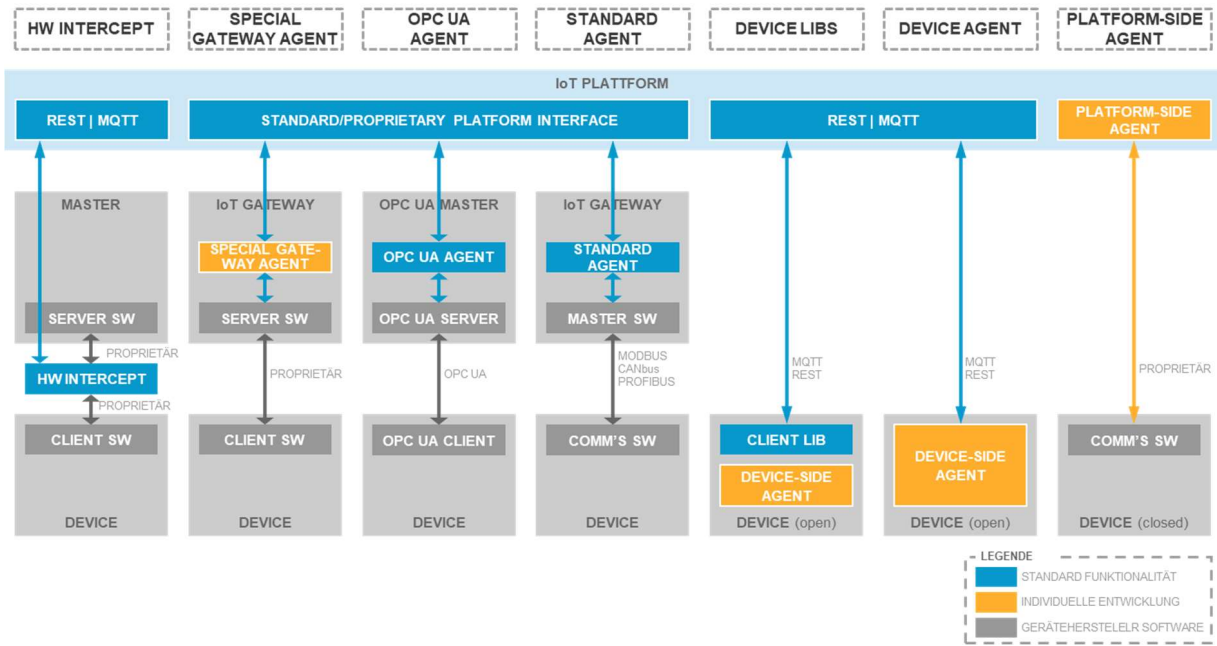

**Abbildung 8.** Architekturmuster zur Anbindung von IoT-Endpunkten (eigene Darstellung)

Zur besseren Skalierbarkeit und rascheren Umsetzung kommen bei etlichen Mustern sogenannte **«IoT-Gateways»** zum Einsatz (vgl. die vier Muster in der linken Hälfte in obiger Abbildung). Deren Aufgabe ist die Konvertierung der physischen und logischen (Kommunikations-)Protokolle der Geräte und IoT Endpunkte in Protokolle, die von der IoT-Plattform unterstützt werden. Die nachstehende Tabelle erläutert die sieben Architekturmuster näher.

**Tabelle 2.** Architekturmuster zur Anbindung von IoT-Endpunkten

| ARCHITEKTUR-MUSTER | ERLÄUTERUNG |
|---|---|
| **PLATFORM-SIDE AGENT** | Für die Anbindung von Geräten, die (Kommunikations-)Protokolle aufweisen, die von der IoT-Plattform nicht *out of the box* unterstützt werden und deren (Kommunikations-)Software durch den Anwender auch nicht verändert oder ergänzt werden kann, besteht oft die Möglichkeit, auf der IoT-Plattform selbst eine geeignete Softwarekomponente zu entwickeln und zu installieren (ein sogenannter «platform-side agent»), die das proprietäre Protokoll des Geräts unterstützt. |

---

[29] *light-weight M2M (protocol).* Ein Client/Server Protokoll zum Device Management.

| **DEVICE AGENT** | Bei Geräten, die die Installation von eigener (Kommunikations-)Software zulassen, kann man eine eigene Softwarekomponente, einen sogenannten «device-side agent» entwickeln, der die internen Daten und Protokolle des Geräts in ein Standardprotokoll der IoT-Plattform übersetzt (hier mit MQTT und REST symbolisiert). |
|---|---|
| **DEVICE LIBS** | Hier liegt dieselbe Situation wie beim Muster DEVICE AGENT vor; mit dem Unterschied, dass der Hersteller der IoT-Plattform für dieses Gerät bereits eine entsprechende Bibliothek («client lib») bereitstellt, die die Entwicklung des «device-side agents» deutlich vereinfacht und beschleunigt. |
| **STANDARD AGENT** | Hier kommt ein sogenannter **«IoT-Gateway»** zum Einsatz, der die physischen (z.B. 2- oder 4-Draht Leitung oder 5G) und logischen Protokolle des IoT-Geräts in Protokolle konvertiert, die von der IoT-Plattform unterstützt werden. Dabei kann es sich um offene Protokollstandards oder auch um proprietäre handeln.<br><br>Bei der aktuellen Option stellt der Hersteller der IoT-Plattform entsprechende Software zur Verfügung (hier «standard agent» genannt), die im Kontext des IoT-Gateways ablaufen und diese Übersetzung vornehmen kann. Der «standard agent» kann dabei zur IoT-Plattform hin (sogenanntes *north-bound interface*) offene oder plattformspezifisch optimierte Protokolle unterstützen. Vgl. dazu das Praxisbeispiel Abschn. 3.2.3.<br><br>IoT-Plattform-Betreiber haben typischerweise Kooperationen mit (zahlreichen) Herstellern von IoT-Gateways, die derart entsprechend unterschiedliche Protokolle (und damit Typen von IoT-Geräten) unterstützen. |
| **OPC UA AGENT** | Hier handelt es sich um eine Ausprägung des STANDARD AGENT- Musters (siehe oben) für das OPC UA Protokoll. |
| **SPECIAL GATEWAY AGENT** | Für den Fall, dass kein geeigneter IoT-Gateway existiert, der *out of the box* das benötigte (Geräte-)Protokoll umwandeln kann, besteht (oft) die Möglichkeit, eine entsprechend individuelle Softwarekomponente zur Protokollumwandlung zu entwickeln (der sogenannte «special gateway agent») und auf einem geeigneten IoT-Gateway zum Einsatz zu bringen. |
| **HW INTERCEPT** | Bei sehr proprietären Gerätefamilien ist es mitunter weder möglich, im Gerät (in der Maschine) selbst, noch in einem «Master» einen «Agent» zur Protokollkonvertierung zwischen Geräten und IoT-Plattform einzusetzen.<br><br>In diesem Fall besteht im Prinzip die Möglichkeit, auf der physischen Ebene (bspw. der 2-Draht Leitung) ein Gerät zum „Abhören" der Übertragungssignale zwischen den «Master» und das Gerät zwischen zu schalten. |

> **Alternative**: Da diese Vorgehensweise sehr aufwändig ist und viel Spezialwissen voraussetzt, installieren viele Organisationen in diesem Fall (sofern zweckmäßig für den IoT-Anwendungsfall) einfach einen entsprechenden neuen Sensor neben oder an der betroffenen Maschine. Diese neue Hardware-Komponente unterstützt dann in aller Regel auch die Protokolle der IoT-Plattform.

## 4.2 «Low Power»-Geräte

Eine oft stillschweigend vorausgesetzte Bedingung für IoT-Projekte betrifft die Versorgung der IoT-Endpunkte mit genügend (i.d.R. elektrischer) Energie zur laufenden Akquisition der entsprechenden (Sensor-)Daten und Kommunikation typischerweise an eine IoT-Plattform. Für IoT-Projekte im industriellen Bereich sind Maschinen und Geräte überwiegend in einer Produktionshalle situiert, wo der ständige Zugang zu einer elektrischen Stromversorgung keine Schwierigkeit darstellt.

Dies ist jedoch für zahlreiche andere Domänen nicht gegeben wie bspw. für Sensoren im Bereich der Umweltanalytik und Landwirtschaft, für Wasserzähler (selbst im verbauten Gebiet), aber auch für Güterwaggons[30], insofern man sie mit Sensoren ausstatten wollte.

Die IoT-Geräte müssen in diesen Fällen ihre Energie aus (elektrischen) Batterien beziehen. Dabei stellt sich jedoch heraus, dass gängige (hier vor allem: drahtlose) Übertragungsprotokolle und -Standards zu wenig Rücksicht auf die begrenzte Energiekapazität der Geräte nehmen. Der Betreiber einer derartigen IoT-Lösung wäre dann gezwungen, die Batterien bei sämtlichen (Hunderten oder Tausenden) IoT-Geräten regelmäßig in kurzen Abständen auszutauschen.

Zur Vermeidung dieser Situation haben sich Niedrigenergiestandards im Bereich der drahtlosen Weitverkehrsnetzwerke (*low power WAN*, LPWAN) herausgebildet, bei denen ein Batterietausch beim IoT-Gerät nur alle 3-10 Jahre vorzunehmen ist.

Hier sind allerdings deutliche Kompromisse zwischen Batterielebensdauer und Übertragungsgeschwindigkeit und Nachrichtengröße einzugehen (vgl. Abbildung 9 mit Daten aus [64]).

---

[30] Im Gegensatz zu Personenzügen, bei denen jeder einzelne Personenwaggon über einen Anschluss an ein Stromnetz verfügt, ist dies bei Güterzügen nur für das Zugfahrzeug („Lokomotive") der Fall. Die Güterwaggons selbst haben keinen Zugang zu elektrischer Energie.

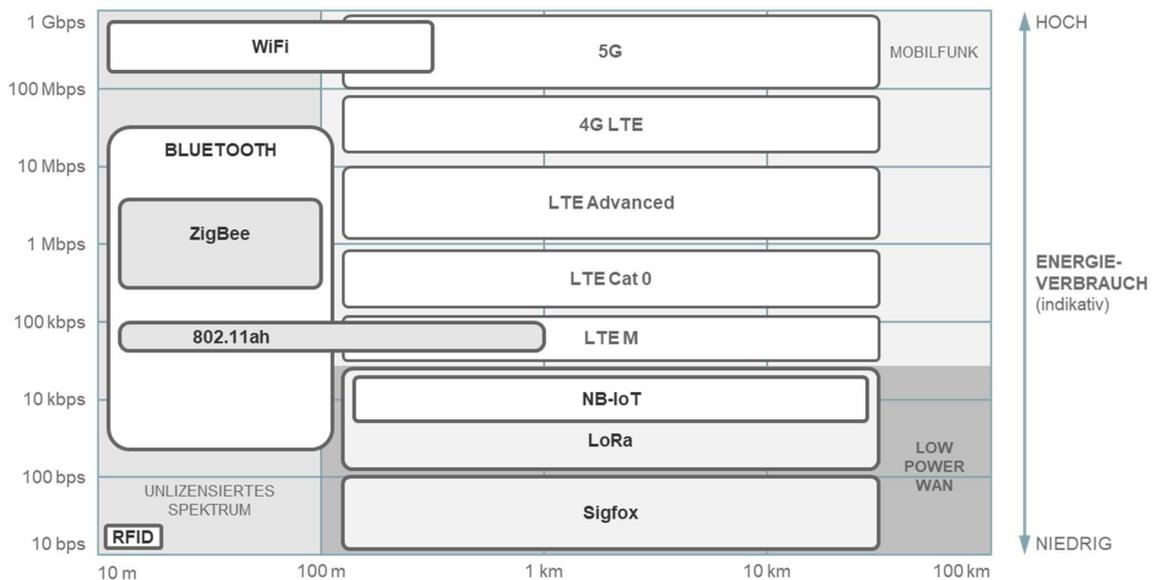

**Abbildung 9.** Reichweite vs. Bandbreite drahtloser Protokolle (eigene Darstellung. Daten aus [64])

So erreicht man bei Einsatz des NB-IoT[31]-Protokolls für Wasserzähler eine durchschnittliche Batterielebensdauer von ca. 5 Jahren, wenn der Wasserzähler pro Tag lediglich eine einzige Nachricht mit weniger als 100 Bytes an die IoT-Plattform sendet. Eine derartige Nachricht umfasst dann bspw. nur die gemessene Durchflussmenge seit der letzten Kommunikation und die durchschnittliche Wassertemperatur.

Bei Güterwaggons optimiert man die Batterielebensdauer der IoT-Geräte, die direkt am Waggon angebracht sind, indem sie bei stehendem Waggon nur alle 12 Stunden eine Nachricht an die IoT-Plattform senden. Setzt sich der Waggon jedoch in Bewegung, wird sofort eine Nachricht versendet, sowie im Zustand der Bewegung regelmäßig alle 15 Minuten.

---

[31] *narrow-band IoT,* ein 4G/5G Standard

# 5  INFORMATIONSALLOKATION UND -DISTRIBUTION

## 5.1  Lambda-Architektur

Die nachstehende Tabelle fasst typische Größenordnungen für Anlagen in der produzierenden Industrie (man denke etwa an eine Automobilwerk) zusammen[32].

**Tabelle 3.** Größenordnung der Komponenten bei IIoT-Projekten

| GRÖSSENKLASSE DER ANLAGE | SENSOREN | STEUER-SYSTEME | IoT-GATEWAYS | DATENSTROM PRO TAG |
|---|---|---|---|---|
| **KLEIN** | < 50.000 | < 2.000 | < 50 | < 200 GB |
| **MITTEL** | 50.000 – 80.000 | 2.000 – 4.000 | 50 – 100 | 200 – 300 GB |
| **GROSS** | > 80.000 | > 4.000 | > 100 | > 300 GB |

Diese Zahlen bedeuten, dass eine IoT-Plattform für eine einzige (!) Produktionsanlage ständig einen Datenstrom von ca. 1,0-3,5 MB/s zu verarbeiten hat. Erweitert man dies auf einen global tätigen Industriekonzern mit ca. 100 Industrieanlagen gemischter Größenklassen so kommt man auf 5-8 Mio. Sensoren, 250-300.000 Steuersysteme und 5-7.000 IoT-Gateways, die einen kontinuierlichen Datenfluss von 15-20 TB pro Tag erzeugen.

Ein System zur Echtzeitanalyse (*real time analytics*) derart hoher Datenströme mit entsprechend niedriger Latenz[33] muss daher einen Datenfluss von ca. 150–250 MB/s bewältigen — ein Unterfangen, das sich mit klassischen (relationalen) Datenbanken nicht bewerkstelligen lässt. Parallel dazu benötigt man dieselben Daten auch für die extensive (und akzeptierterweise länger dauernder) Analyse im Kontext von Big Data-Verfahren.

Als eine Lösung dieser Anforderungen hat sich in den letzten Jahren der Einsatz der sogenannten **Lambda-Architektur** in vielen Fällen bewährt [65][66][67][68]. Das wesentlichen Charakteristikum dieses (Enterprise IT) Architektur-Musters besteht in der frühen Trennung[34] der Analyse von Echtzeitdaten (*data in motion*) und allen anderen Daten (*data at rest*). Erstere werden in der sogenannten «Geschwindigkeits-Schicht» (*speed layer*) analysiert, während letztere in einer Batch-Schicht (*batch layer*) persistiert und analysiert werden (vgl. das Datenflussdiagramm in Abbildung 10.

---

[32] Projekterfahrungen des Autors
[33] wo also das Ergebnis der Analyse sehr rasch, sozusagen *on the fly* (in der Größenordnung von Sekunden), verfügbar sein muss
[34] der sogenannte *Lambda fork* (eigene Bezeichnung). In der obigen Architektur befindet sich diese Aufspaltung (*fork*) *nach* der IoT-Plattform; für extrem hohe Datenströme wird dieser Verzweigungspunkt mitunter auch *vor* die IoT-Plattform bzw. an einen geeigneten Punkt im Edge-Kontinuum gelegt.

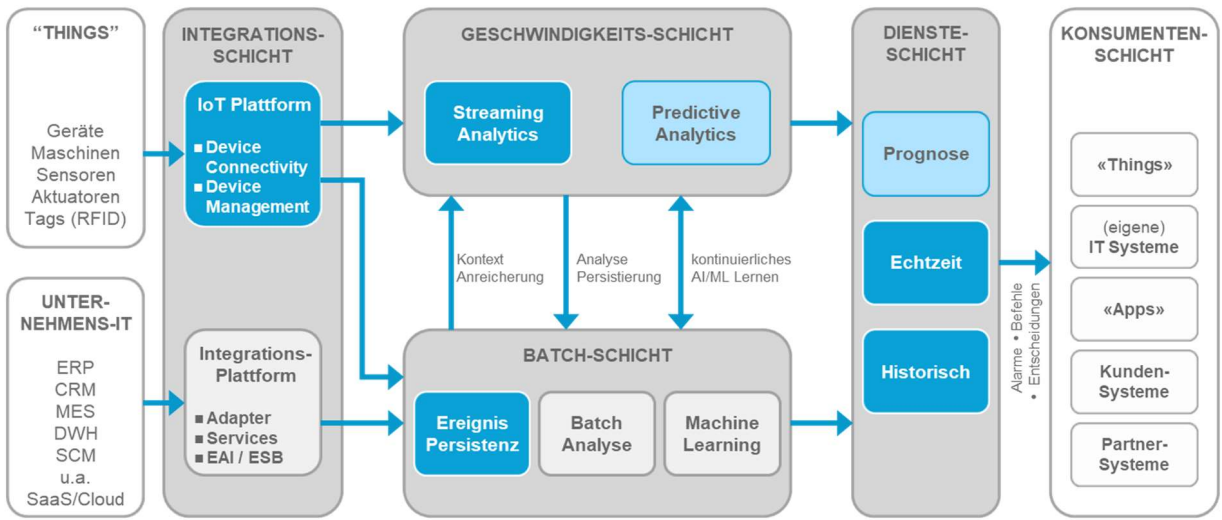

**Abbildung 10.** Lambda-Architektur für IT/OT-Integration (eigene Darstellung)

Die Ergebnisse beider Verarbeitungsschichten werden dann über die sogenannte «Dienste-Schicht» (*service layer*) direkt für den Endbenutzer oder für nachfolgende Systeme aufbereitet bzw. weitergeleitet[35]. Diesen beiden Verarbeitungsschichten wird in der Regel eine Integrations-Schicht vorgelagert, die für den Bereich der IoT-Endpunkte typischerweise durch Funktionen der IoT-Plattform, für die Systeme der „klassischen" Unternehmens-IT durch Integrationsplattformen aus der Domäne des *Enterprise Application Integration* (EAI) bzw. durch einen Enterprise Service Bus (ESB) realisiert wird. In Blau gefärbte Elemente in der obigen Abbildung reflektieren dabei Funktionen, die typischerweise durch IoT-Plattformen realisiert werden.

Wie jeder Architektur ist auch der Lambda Architektur ein bestimmter *trade off* zu eigen, bei dem die Architektur explizit oder auch implizit im Verhältnis zu den Anforderungen bestimmte Abwägungen und Entscheidungen trifft. In diesem Fall betrifft das die Aufgabe, die Geschäftsregeln und Daten der Geschwindigkeits-Schicht mit der Batch-Schicht zu synchronisieren. Erfolgen die Änderungen und damit der Koordinierungsaufwand zwischen den beiden Schichten mit zu hoher Frequenz (etwa minütlich), so ist das in der Lambda Architektur nicht mehr effizient implementierbar, sondern hat zur Herausbildung alternativer Architekturmuster geführt (wie bspw. der Kappa-Architektur), die etwa ausschließlich auf die Verarbeitung von Datenströmen oder von Mikro-Batches optimiert sind [69].

## 5.2   API-Management

Der unternehmensübergreifende Austausch digitaler Daten (*electronic data interchange,* EDI) im sogenannten *extended enterprise* hat im gewerblich-industriellen Bereich seit mehr als drei Dekaden eine erfolgreiche Tradition [30]. Anfangs war dies auf Grund der Komplexität der *business-to-business* (B2B) Standards auf große Unternehmen mit entsprechend hochvolumigem Datenaustausch beschränkt, die sich die notwendige IT-Infrastruktur oft in der Form von EDI-

---

[35] Dies führt eigentlich dazu, dass die Systeme auf der rechten Seite der Abbildung teilweise mit den Ursprungssystemen auf der linken Seite überlappen und das Architekturdiagramm derart mit einer Torus-Topologie versehen.

Konvertern leisten konnten. Die synchrone Weiterentwicklung von service-orientierten Architekturen (SOA) und dem World Wide Web (bspw. durch den HTTP/REST- Standard aus dem Jahr 2000 bzw. Web-APIs, nicht zuletzt für mobile Endgeräte wie *smart phones* [70][71]), hat nunmehr zur Konvergenz beider Entwicklungsstränge in Form der **API-Ökonomie** (*API economy*) geführt [72][73].

Dabei werden unternehmensinterne (informationstechnische) Fähigkeiten und Funktionen über programmatische Schnittstellen — den sogenannten *application programming interfaces* — exponiert und Dritten zur Verfügung gestellt. Dadurch werden diese Organisationen (genauer: ihre jeweiligen Software-Entwickler) in die Lage versetzt, diese Funktionen ihrerseits in Informationssysteme und Applikationen — inklusive *Apps* für mobile Endgeräte — zu integrieren [71][74]. Dieses durchaus auch entgeltliche Abgreifen von elektronischen Funktionen im *extended enterprise* konstituiert dann die API-Ökonomie.

Die damit im unternehmensübergreifenden Kontext erzielbaren ökonomischen Vorteile sind gut dokumentiert [74][75] und wurden darüber hinaus sogar (indirekt) von der EU etwa im Rahmen ihrer (zweiten) Zahlungsdiensterichtlinie[36] aufgegriffen, die eine derartige Öffnung von vormals rein internen Unternehmensdaten und -funktionen gegenüber (vertrauenswürdigen) Dritten im Bankenkontext obligatorisch macht.

Primär aus Sicherheitsgründen, sekundär aber zur Bereitstellung der im Rahmen des sogenannten **API-Managements** [76] notwendig werdenden Zusatzfunktionen bündelt man alle diesbezüglichen Fähigkeiten in einer **API-Plattform** [71][77](vgl. Abbildung 11).

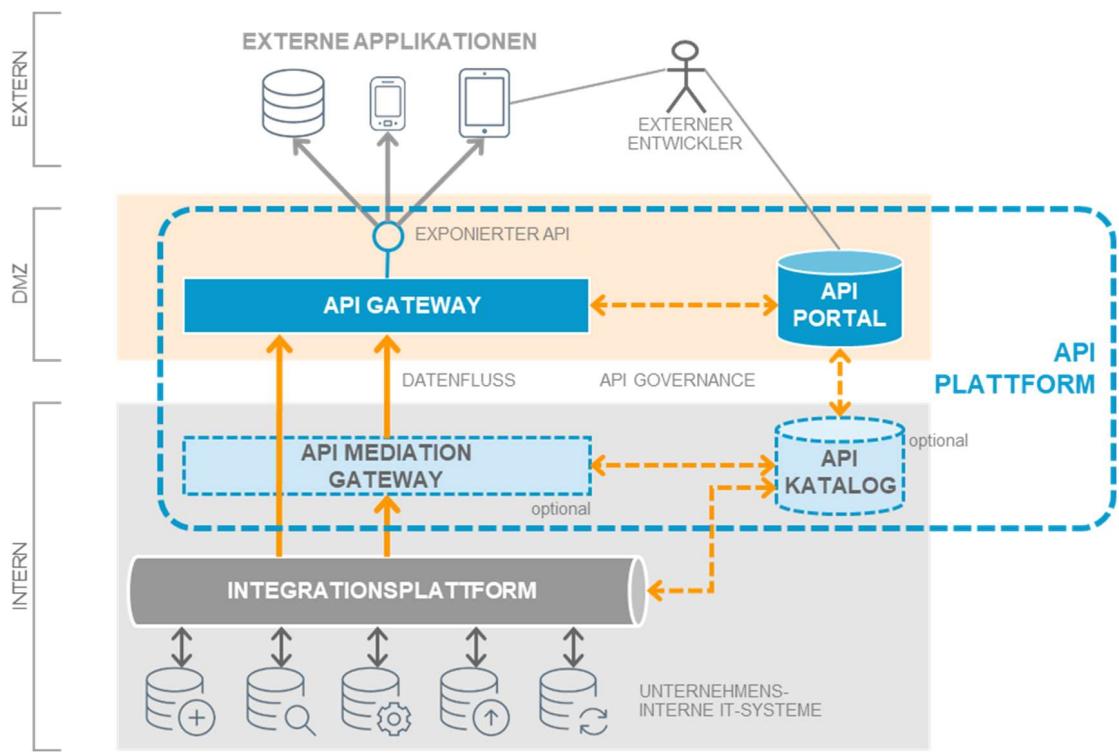

**Abbildung 11.** Komponentenarchitektur im API-Management (eigene Darstellung)

---

[36] Richtlinie (EU) 2015/2366 vom 25. Nov. 2015 (*Payments Service Directive 2*, PSD2)

Die beiden wichtigsten Komponenten einer API-Plattform sind der **API-Gateway** und das **API-Portal**.

Der API-Gateway ist in der demilitarisierten Zone (DMZ) lokalisiert und stellt die (internen) Dienste und Funktionen technisch anderen (externen) Systemen über vorab definierte Schnittstellen (Protokoll- und Datenebene) zur Verfügung. Gleichzeitig schirmt es im Sinne eines *Firewalls* die internen Systeme vor den externen Zugriffen ab (i.d.R. über die Implementierung des *reverse proxy* Musters), schützt vor Angriffen (wie *denial of service*-Attacken) und stellt sicher, dass nur berechtigte externe Applikationen auf die internen Funktionen zugreifen dürfen (Authentifizierung und Autorisierung).

Das API-Portal bündelt alle Informationen, die ein Software-Entwickler benötigt, um die via API-Gateway angebotenen APIs in seine eigene, je individuellen Applikationen und Systeme integrieren zu können. Neben der reinen Dokumentation der APIs finden sich in API-Portalen typischerweise auch Erläuterungen und Daten zum Testen der APIs sowie zum Erlangen der in aller Regel notwendigen sicherheitstechnischen Autorisationen.

In komplexeren IT-Infrastrukturen verlagert man gerne bestimmte Funktionen aus dem API-Gateway in einen im internen Netzwerk (Intranet) befindlichen API-Mediation-Gateway, um noch flexibler zu sein[37]. Dies umfasst Funktionalitäten wie bspw. das Routing von API-Anfragen, die Anpassung (Mediation) von externen APIs an die Formate der internen APIs oder auch die Implementierung von API-spezifischen Richtlinien (*policies*). Insoweit APIs auch im unternehmensinternen Kontext eine wesentliche Rolle spielen [78], wird häufig ein „internes" API-Portal eingesetzt, in der obigen Abbildung API-Katalog genannt. Dieses stellt dann weiterführende Funktionen, wie eine Abhängigkeitsanalyse der (internen) APIs untereinander oder die Steuerung der Aktivitäten des gesamten **API-Lebenszyklus,** zur Verfügung.

Ein Aspekt des Transformationscharakters von Industrie 4.0 und IIoT kann sehr gut durch die Strategie der **„Servifizierung"** klassischer physischer Produkte beschreiben werden. Im Sinne des Beispiels der Kompressoren aus Abschn. 3.2.3 wäre das der Übergang vom Verkaufen oder Verleasen der physischen Maschinen in Richtung der Bereitstellung einer Dienstleistung (*service*) á la „komprimierte Luft *as a service*". Solche innovative Geschäftsmodelle erfordern u.a. auch die Bereitstellung des entsprechenden APIs, durch die der Endkunde die Hauptparameter der aktuell bereit gestellten Druckluft (Druck, Temperatur, Dichte, aktueller Stromverbrauch u.a.m.) abgreifen kann, damit die die Druckluft konsumierende Maschine effizient und effektiv (bspw. innerhalb bestimmter Qualitätsgrenzen) läuft. Damit bekommt dem API-Management eine Schlüsselrolle bei der Umsetzung innovativer, IIoT-getriebener Geschäftsmodellen zu.

Besonders (aber nicht ausschließlicherweise) bei der Monetarisierung der extern angebotenen APIs mutiert der oder die für einen API (in der Praxis: eine größere kohärente Menge an APIs) Verantwortliche dann zu einem (API-)Produkt Manager. Dabei zeigt sich, dass im Allgemeinen die Menge an APIs in Abhängigkeit von ihrer Stabilität bzw. Änderungsfrequenz in unterschiedliche Schichten gegliedert werden kann [77], von User (*experience*) APIs über Prozess-APIs zu System-APIs. Dementsprechend verlagern sich auch die Verantwortungen von der Fachabteilung (*line of business*, LoB) für APIs nahe am Endbenutzer zur zentralen IT für system- und applikationsnahe APIs.

---

[37] hierbei handelt es sich um die Anwendung der klassischen Architekturregel "*separation of concerns*"

# 6 DISKUSSION

**IoT-Plattformen** stehen zu Recht im Mittelpunkt der Lösungsarchitektur von Projekten zur Realisierung von IoT- bzw. IIoT-Vorhaben, da sie als Middleware die Komplexität der Anbindung und des Managements der intrinsisch heterogenen Geräte und Maschinen und sonstigen IoT-Endpunkte gegenüber IoT-Frameworks oder anderen Ansätzen deutlich vereinfachen bzw. effizienter machen.

Unsere Ausführungen legen aber nahe, dass sachlich-technische Zwänge dazu führen, dass die Verbindung zwischen dem *thing* im IoT und der IoT-Plattform in der Cloud in vielen Fällen zum **Edge-Kontinuum** erweitert werden muss. Dabei müssen Untermengen der Funktionalitäten der Cloud-basierten IoT-Plattform auf IoT-Gateways und andere Systeme näher an die IoT-Endpunkte gebracht werden, um hohe nicht-funktionale Anforderungen in den Bereichen Latenz, Bandbreite, Rechenkapazität sowie Sicherheit und Vertraulichkeit ökonomisch gewährleisten zu können.

Ähnliches gilt für den Bereich der **IT/OT-Integration**, der inhaltlich die informationstechnische Koppelung der klassischen Automatisierungstechnik an die klassische IT-Landschaft eines Unternehmens mit seinen ERP, CRM und zahlreichen anderen Unternehmensapplikationen bezweckt. Auf Grund der (zur Genüge bekannten) Komplexität der Aufgabe der *enterprise application integration* (EAI) erfordert dies die Ergänzung der IoT-Plattform mit einer entsprechenden **Integrationsplattform**, bspw. einer *integration PaaS*-Lösung, einer EAI-Plattform oder einem *enterprise service bus* (ESB) (siehe Abbildung 12).

Die Konvergenz von Web-Technologien und service-orientierten Ansätzen hat mittlerweile zur Ausbildung der **API-Ökonomie** geführt, in der Unternehmen im Rahmen des sogenannten *extended enterprise* Funktionen ihrer internen (!) IT-Systeme und Applikationen über entsprechende programmatische Schnittstellen (APIs) Dritten mitunter auch entgeltlich zur Verfügung stellen und derart gemeinsam Effizienzgewinne, etwa in der Logistik oder Supply Chain, realisieren können. Die Funktionen des dadurch bedingten API-Managements und insbesondere sicherheitsrelevante Anforderungen bedingen diesfalls den Einsatz einer weiteren Plattform, der **API-Plattform**.

Man bemerke, dass in Abbildung 12 das Konzept der Integrationsplattform (iPaaS[38]/EAI/ESB) durch den Betrieb eines sogenannten «Integration Agent» im Bereich des externen Dritten ausgedehnt wurde. Man bedient sich dieses Architekturmusters, wenn man *end-to-end*-Kommunikation mit einem Partner herstellen möchte, diesem aber die technischen Mittel zur Integration mit seinen eigenen IT-Systemen fehlen.

---

[38] *integration PaaS*

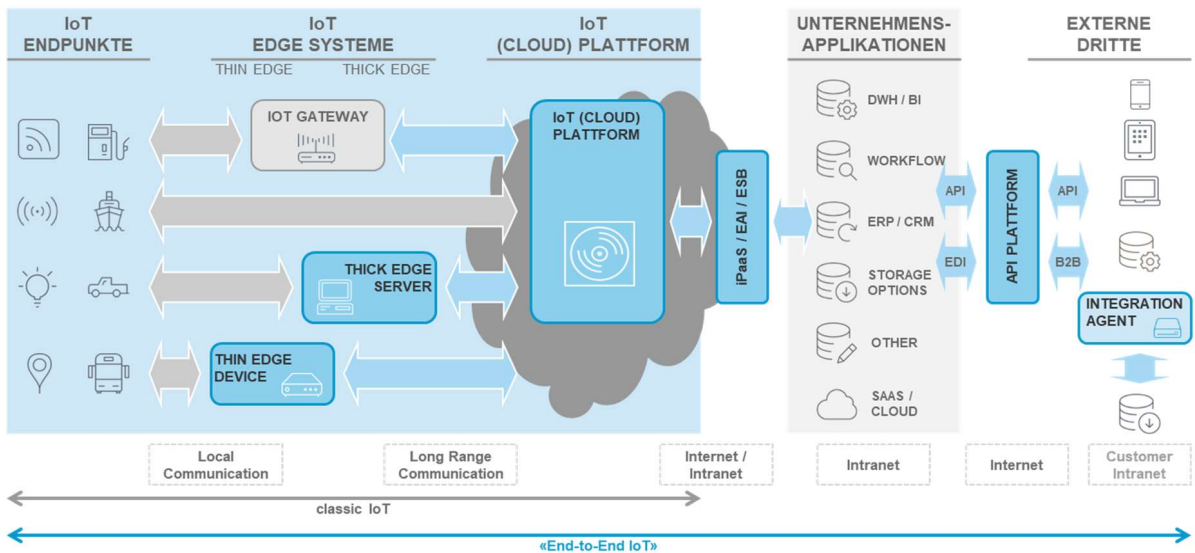

**Abbildung 12.** *End-to-end*-Architektur im IIoT (eigene Darstellung)

Es versteht sich von selbst, dass die Realisierung von *End-to-End*-IoT ein langfristig-strategisches Vorhaben ist, bei dem Transformationspotenzial und Innovationskraft der einzelnen Initiativen mit der Komplexität der entsprechenden Umsetzungsprojekte korrelieren. Daher haben sich in der Unternehmenspraxis Vorgehensmodelle entwickelt, die drei je unterschiedliche Phasen oder Stufen unterschiedlichen Wertbeitrags erkennen lassen (siehe Abbildung 13). Diese Stufen können dabei durchaus im Sinne eines Reifegradmodells betrachtet werden, bei dem das Erreichen einer bestimmten Maturität die Voraussetzung für das Erlangen der nächsthöheren Stufe darstellt.

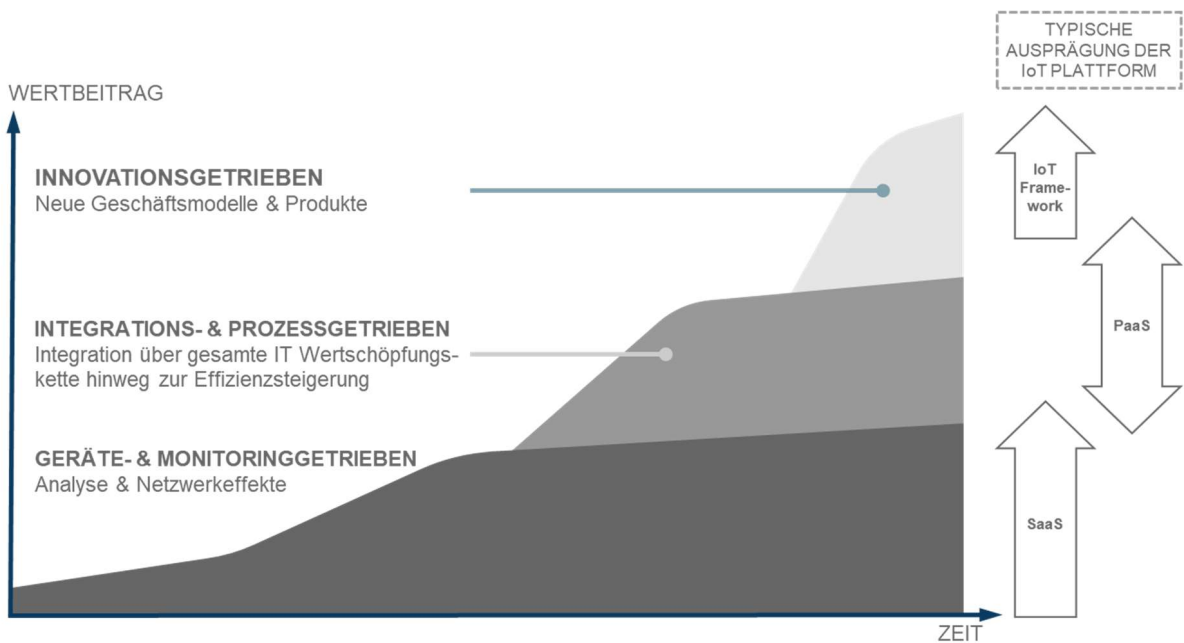

**Abbildung 13.** Phasenmodell der Umsetzung der IoT-Strategie (eigene Darstellung)

In der initialen Phase werden die entsprechenden Geräte, Maschinen und andere IoT-Endpunkte erstmals an eine geeignete IoT-Plattform angeschlossen, um überhaupt Zugang zu deren Daten herzustellen. Die — oftmals zunächst rein visuelle — Darstellung und Analyse dieser Daten über physische Grenzen (bspw. einzelne Stationen) der Produktionsinfrastruktur hinweg erlauben nach dem Zustandsmonitoring der IoT-Endpunkte auch ein erstes Korrelieren und damit die Realisierung von Netzwerkeffekten. In diesem Stadium werden viele Anforderungen an eine IoT-Plattform von Standardangeboten der Hersteller *out of the box* erfüllt. Dies führt zu einer Prävalenz des SaaS-Modells für den Bezug dieser Funktionalitäten.

Diese Effekte erreichen jedoch nach einer bestimmten Zeit durch den abnehmenden Grenznutzen ein gewisses Nutzenplateau. Bereits in der nächsten Phase geht man daher über eine „klassische", an der IoT-Plattform endende Konzeption von IoT/IIoT hinaus, indem man die IoT-Plattform und ihre Daten an die übrigen Unternehmenssysteme und -applikationen koppelt (typischerweise über Integrationsplattformen). Dadurch lassen sich über die applikations- und plattformübergreifende Automatisierung von Unternehmensprozessen entsprechende Effizienzgewinne erzielen. Dies kann auch im Rahmen der aktiven Partizipation an der API Ökonomie die eigenen Unternehmensgrenze durch Einsatz einer API-Plattform in Richtung des Unternehmensumfelds des *extended enterprise* überschreiten. Die in dieser Stufe deutlich steigenden Individualisierungs- und Adaptionsanforderungen an die involvierten Plattformen können hier oft durch PaaS-Leistungen abgedeckt werden.

Beiden Stufen gemeinsam ist oftmals die Tatsache, dass sich die Projekte innerhalb der Limitationen bestehender Produkte, Leistungen und Geschäftsmodelle bewegen. Die erfolgreiche Bewältigung dieser ersten Phasen stellt aber nahezu paradigmatisch die technisch-organisatorischen Voraussetzungen für das Transzendieren dieser Grenzen durch entsprechende Geschäftsmodellinnovation dar. Man denke etwa an die **„Servifizierung"** klassischer physischer Produkte (siehe dazu das Beispiel „komprimierte Luft *as a service*" in Abschn. 5.2). Durch den Novitätscharakter derartiger Initiativen trifft man hier bei der technischen Umsetzung öfter auf unternehmensindividuelle IoT-Frameworks, die allerdings bestehende (IoT- und andere) Plattformen mitintegrieren bzw. geradezu auf diesen aufbauen und diese nur in den nicht abgedeckten Bereichen funktional augmentieren.

Mit und auf dieser dritten und letzten Stufe kann dann der anfangs angesprochene Innovations- und Transformationsprozess von Industrie 4.0 sein gesamtes Potenzial ausschöpfen und damit tatsächlich — naturgemäß abhängig von der Geschwindigkeit der Adoption in der breiten Fläche — eine Entwicklung bewirken, die man *ex post* vielleicht einmal als vierte industrielle Revolution bezeichnen wird. Ganz im Gegensatz zum diskontinuierlichen Charakter einer (echten) Revolution ist bei dem in diesem Beitrag skizzierten (technischen) Vorgehen kontinuierlicher, auf jeder Stufe und in jeder Phase deutlicher und nachhaltiger Nutzen erzielbar.

## LITERATURVERZEICHNIS

[1]   Al-Fuqaha A. et al.(2015) Internet of Things: A survey on enabling technologies, protocols, and applications. IEEE Commun. Surveys Tuts. 17(4), S.2347-2376.

[2]   Hansen H.R., Mendling J. und Neumann G. (2019) Wirtschaftsinformatik. 12. Auflage, Berlin: Walter de Gruyter.

[3]   Heimann B. et al. (2016) Mechatronik. Komponenten - Methoden - Beispiele. München: Carl Hanser, S. 29ff.

[4]   Sendler, U. (2013) Industrie 4.0–Beherrschung der industriellen Komplexität mit SysLM (Systems lifecycle management). In Sendler U. (Hrsg.) Industrie 4.0. Berlin: Springer, S.1-19.

[5]   Kim J et al. (2014) M2M Service Platforms: Survey, Issues, and Enabling Technologies. IEEE Comm. Surv & Tut. 16(1), S.61-76.

[6]   Vaillant S. (2020). Private Mitteilung (01. November 2020).

[7]   Jeschke, S., Brecher, C., Meisen, T., Özdemir, D. and Eschert, T. (2017). Industrial Internet of Things and Cyber Manufacturing Systems. In *Industrial Internet of Things* (S. 3-19). Springer, Cham.

[8]   Liao Y. et al. (2018) Industrial Internet of Things: A Systematic Literature Review and Insights. IEEE IoT Journal 5(6), S.4515-4525.

[9]   Fraile F. et al. (2018) Trustworthy Industrial IoT Gateways for Interoperability Platforms and Ecosystems. IEEE IoT Journal 5(6) S.4506-4514.

[10]  Hänisch W. (2017): Grundlagen Industrie 4.0. In: Andelfinger V.P. und Hänisch T. (Hrsg) "Industrie 4.0. Wie cyber-phyische Systeme die Arbeitswelt verändern". Wiesbaden: Springer, S.9-32.

[11]  Oztemel, E. and Gursev, S., 2020. A Taxonomy of Industry 4.0 and Related Technologies. In *Industry 4.0-Current Status and Future Trends*. IntechOpen

[12]  Platform Industrie 4.0 (2015) Umsetzungsstrategie Industrie 4.0. Ergebnisbericht der Plattform Industrie 4.0. April 2015, S.8

[13]  Qi, Q., & Tao, F. (2018) Digital twin and big data towards smart manufacturing and industry 4.0: 360 degree comparison. IEEE Access, 6, S.3585-3593.

[14]  Middleton P., Sharpington K. & Eschinger C. (2019) Forecast: Internet of Things, Endpoints and Communications, Worldwide, 2019-2029. G00719959, Gartner Inc.

[15]  MacGillivray C. & Reinsel D. (2019) Worldwide Global DataSphere IoT Device and Data Forecast, 2019–2023. IDC Market Forecast.

[16]  Schmitz C. et al. (2019) Industry 4.0. Capturing value at scale in discrete manufacturing. McKinsey. S.6.

[17]  Voß S. & Gutenschwager K. (2001) Informationsmanagement. Berlin:Springer.

[18]  Olsen, T. L., & Tomlin, B. (2020) Industry 4.0: opportunities and challenges for operations management. Manufacturing & Service Operations Management, 22(1), S.113-122.

[19]  Slack N., Chambers S. & Johnston R. (2001) Operations Management. Essex (England):Pearson Education.

[20]  Strnadl, C.F. (2006) Aligning business and it: The process-driven architecture model. Information Systems Management, 23(4), S.67-77.

[21]   Manouvrier, B. and Ménard, L., 2010. Application Integration: EAI B2B BPM and SOA (Vol. 130). John Wiley & Sons.

[22]   Ferstl O.K. & Sinz E.J. (2008) Grundlagen der Wirtschaftsinformatik. München:Oldenbourg

[23]   Hansen H.R. & Neumann G. (2005) Wirtschaftsinformatik 1. Grundlagen und Anwendungen. 9. Aufl. Lucius & Lucius: UTB 2699.

[24]   Völter, M., Kircher, M. and Zdun, U. (2013) Remoting patterns: foundations of enterprise, internet and realtime distributed object middleware. John Wiley & Sons.

[25]   Ngu, A.H., Gutierrez, M., Metsis, V., Nepal, S. and Sheng, Q.Z. (2017) IoT middleware: A survey on issues and enabling technologies. IEEE IoT Journal, 4(1), S.1-20.

[26]   Razzaque, M.A., Milojevic-Jevric, M., Palade, A. and Clarke, S. (2016). Middleware for internet of things: a survey. IEEE IoT Journal, 3(1), S.70-95.

[27]   Eclipse (2016). The Three Software Stacks Required for IoT Architectures. IoT software requirements and how to implement them using open source technology. Eclipse Foundation/iot.eclipse.org

[28]   da Cruz M.A.A. et al. (2018) A Reference Model for Internet of Things Middleware. In: IEEE IoT Journal 5(2), S.871-883.

[29]   Machnation (2016). Build, acquire, and buy: An AEP analysis. https://www.machnation.com/2016/09/30/build-acquire-buy-aep-analysis/ Zugegriffen: 13. Oktober 2020

[30]   Jagdev, H. S., & Browne, J. (1998). The extended enterprise - A context for manufacturing. Production Planning & Control, 9(3), S.216-229.

[31]   Mineraud, J., Mazhelis, O., Su, X. and Tarkoma, S., 2016. A gap analysis of Internet-of-Things platforms. Computer Communications, 89, S.5-16.

[32]   Vogel B. et al. (2020) What Is an Open IoT Platform? Insights from a Systematic Mapping Study. Future Internet 12(73), S.1-19.

[33]   Song, X., Hwong, B. and Ros, J. (2011) Lessons from developing nonfunctional requirements for a software platform. IEEE Software, 29(2), S.74-80.

[34]   Hilkert, D., Benlian, A., Sarstedt, M., & Hess, T. (2011). Perceived software platform openness: the scale and its impact on developer satisfaction. In: International Conference on Information Systems 2011 (ICIS 2011), S.3188-3207.

[35]   Santana E.F.Z. et al. (2017) Software Platforms for Smart Cities: Concepts, Requirements, Challenges, and a Unified Reference Architecture. ACM Comput. Surv. 50, 6, Article 78 (January 2018), S. 1-37.

[36]   Mijuscovic A. et al. (2020) Comparing Apples and Oranges in IoT context: A deep dive into methods for comparing IoT platforms. IEEE IoT Journal, doi: 10.1109/JIOT.2020.3016921, to appear.

[37]   Falck, O., Koenen, J. (2020): Industrielle Datenwirtschaft - B2B Plattformen. ifo Studie im Auftrag des BDI.

[38]   BDI (2020): Deutsche digitale B2B-Plattformen. BDI Publikationsnr. 0102.

[39]   Asemani, M., Abdollahei, F., Jabbari, F. (2019) Understanding IoT platforms: towards a comprehensive definition and main characteristic description. In 2019 5th International Conference on Web Research (ICWR), S.172-177.

[40]   Hoffmann J.P., Heimes P., Senel S. (2019) IoT Platforms for the Internet of Production. In: IEEE IoT Journal 6(3), S.4098-4105.

[41]   Ullah M. et al. (2020): Twenty-One Key Factors to Choose an IoT Platform: Theoretical Framework and Its Applications. IEEE IoT Journal 7(10), S.10111-10118.

[42]   Weyrich M. and Ebert C. (2016) Reference Architecture for the IoT. IEEE Software, 33(1), S.112-116.

[43]   Shi W, Dustdar S. (2020) The Promise of Edge Computing. In: IEEE Computer 49(5), S.78-81.

[44]   Pan, J. and McElhannon, J. (2018) Future edge cloud and edge computing for internet of things applications. IEEE IoT Journal, 5(1), S.439-449.

[45]   Taleb, T. et al. (2017) On multi-access edge computing: A survey of the emerging 5G network edge

cloud architecture and orchestration. IEEE Communications Surveys & Tutorials, 19(3), S.1657-1681.

[46] Premsankar, G., Di Francesco, M. and Taleb, T. (2018) Edge computing for the Internet of Things: A case study. IEEE IoT Journal, 5(2), S.1275-1284.

[47] Pham, Q.V.et al. (2020). A survey of multi-access edge computing in 5G and beyond: Fundamentals, technology integration, and state-of-the-art. IEEE Access, 8, S.116974-117017.

[48] Hu P. et al. (2020) Software-Defined Edge Computing (SDEC): Principle, Open IoT System Architecture, Applications, and Challenges. IEEE IoT Journal, 7(7) S.5934-5945.

[49] Cui L. et al. (2020) A Decentralized and Trusted Edge Computing Platform for Internet of Things. IEEE IoT Journal, 7(5), S.3910-3922.

[50] Johnson, W., Sparks, K., Daly, B., et al. (2019). 5G Edge Computing White Paper. FCC Technological Advisory Council, 5G IOT Working Group. https://transition.fcc.gov/bureaus/oet/tac/tacdocs/reports/2018/5G-Edge-Computing-Whitepaper-v6-Final.pdf. Zugegriffen am: 27.10.2020

[51] ETSI (2019) Multi-access edge computing: Terminology. ETSI GS MEC 001 V2.1.1 (2019-01).

[52] Chiang, M. and Zhang, T. (2016) Fog and IoT: An overview of research opportunities. IEEE Internet of Things Journal, 3(6), S.854-864.

[53] Naha, R.K. et al. (2018) Fog Computing: Survey of trends, architectures, requirements, and research directions. IEEE Access, 6, S.47980-48009.

[54] Omoniwa B. et al. (2019) Fog/Edge Computing-Based IoT (FECIoT): Architecture, Applications, and Research Issues. IEEE IoT Journal 6(3), S.4118-4149.

[55] Mouradian, C. et al. (2017) A comprehensive survey on fog computing: State-of-the-art and research challenges. IEEE Communications Surveys & Tutorials, 20(1), S.416-464.

[56] Iorga M. et al. (2018) Fog Computing Conceptual Model. Recommendations of the National Institute of Standards and Technology. NIST Special Publication 500-325.

[57] Bittmann T. (2020) Why and How I&O Should Lead Edge Computing. Gartner Report G00467223.

[58] Varghese, B., Reano, C., & Silla, F. (2018) Accelerator virtualization in fog computing: Moving from the cloud to the edge. IEEE Cloud Computing, 5(6), S.28-37.

[59] Silva, D., & Sofia, R. C. (2020) A Discussion on Context-awareness to BetterSupport the IoT Cloud/Edge Continuum. arXiv preprint arXiv:2010.04563.

[60] Spanopoulos-Karalexidis, M. et al. (2020) Simulating Across the Cloud-to-Edge Continuum. In "Managing Distributed Cloud Applications and Infrastructure" (pp. 93-115). Palgrave Macmillan, Cham.

[61] Domaschka, J. et al. (2020) Towards an Architecture for Reliable Capacity Provisioning for Distributed Clouds. In "Managing Distributed Cloud Applications and Infrastructure" (S. 1-25). Palgrave Macmillan, Cham.

[62] Abts D. & Mülder W. (Hrsg.) (2010) Masterstudium Wirtschaftsinformatik. Wiesbaden:Vieweg+Teubner.

[63] Adamos (2020) ADAMOS IIoT. PaaS- und SaaS-basierte Funktionalitäten für den Aufbau von Apps. https://www.adamos.com/loesungen/adamos-iiot/adamos-iiot-funktionalitaeten. Zugegriffen: 23.10.2020.

[64] Alsén D., Patel M. and Shangkuan J. (2017) The future of connectivity. Enabling the Internet of Things. McKinsey High Tech Publication, December 2017.

[65] Marz N. (2011) How to Beat the CAP Theorem. In: "Thoughts from the Red Planet" Blog, 13 Oct. 2011; http://nathanmarz.com/blog/how-to-beat-the-captheorem.html. Besucht am:28.10.2020

[66] Kroß, J., Brunnert, A., Prehofer, C., Runkler, T.A. and Krcmar, H. (2015) Stream processing on demand for lambda architectures. In "European Workshop on Performance Engineering" S.243-257. Springer, Cham.

[67] Hausenblas, M. and Bijnens, N. (2015). Lambda architecture. http://lambda-architecture.net/. Luettu, 6, S.2014.

[68] Kiran, M., Murphy, P., Monga, I., Dugan, J. and Baveja, S.S. (2015) Lambda architecture for cost-effective batch and speed big data processing. In "2015 IEEE International Conference on Big Data (Big

Data)" S.2785-2792.

[69]    Lin J (2017) The Lambda and the Kappa. IEEE Internet Computing 21(5), S.60-66.

[70]    Fremantle, P., Kopecký, J. and Aziz, B. (2015). Web API management meets the Internet of Things. In European Semantic Web Conference (S. 367-375). Springer, Cham.

[71]    De B. (2017) API Management. Berkeley, CA: Apress.

[72]    Vukovic M. et al. (2016) Riding and Thriving on the API Hype Cycle. Guidelines for the enterprise. Communications oft he ACM 59(3), S.35-37.

[73]    Tan, W., Fan, Y., Ghoneim, A., Hossain, M.A. and Dustdar, S., (2016) From the service-oriented architecture to the web API economy. IEEE Internet Computing, 20(4), S.64-68.

[74]    Lindman J. et al. (2020) Emerging Perspectives of Application Programming Interface Strategy: A Framework to Respond to Business Concerns. IEEE Software, 37(2), S.52-59

[75]    Evans, P.C. and Basole, R.C. (2016) Revealing the API ecosystem and enterprise strategy via visual analytics. Communications of the ACM, 59(2), S.26-28.

[76]    Mathijssen, M., Overeem, M. and Jansen, S. (2020) Identification of Practices and Capabilities in API Management: A Systematic Literature Review. arXiv preprint arXiv:2006.10481.

[77]    Weir, L. (2019) Enterprise API Management: Design and deliver valuable business APIs. Packt Publishing Ltd.

[78]    Andreo, S. and Bosch, J. (2019) API Management Challenges in Ecosystems. In "International Conference on Software Business", S.86-93. Springer, Cham.

# ABKÜRZUNGSVERZEICHNIS

Die Fachsprache der IT – wie auch die anderer hochspezialisierter Berufsdomänen (man denke nur an Medizin oder das Rechtswesen — ist überreich an mitunter kryptischen Akronymen und Abkürzungen. Das nachstehende Abkürzungsverzeichnis möchte und kann zwar kein Glossar ersetzen, aber es mag dennoch dem Leser bei der Dekodierung der in diesem Beitrag verwendeten Abkürzungen helfen.

**Tabelle 4.** Abkürzungsverzeichnis

| Abbreviation | Full text |
|---|---|
| µC | Microcontroller |
| 3G | Mobilfunkt 3.Generation (v.a. UMTS) |
| 4G | Mobilfunk der 4. Generation (LTE) |
| 5G | Mobilfunk der 5. Generation (NR) |
| AG | Aktiengesellschaft |
| AI | *artificial intelligence* |
| API | *applications programming interface* |
| AT | Automatisierungstechnologie (= OT im Englischen) |
| B2B | *business to business* |
| BI | *business intelligence* |
| BPMS | *business process management system* |
| C2E | *cloud-to-edge* |
| CAN | *controller area network* |
| CoAP | *constrained application protocol* |
| Comm's | *communications* |
| CPS | *cyber-physical system* |
| CPU | *central processing unit* |
| CRM | *customer relationship* (System) |
| CSP | *communications services provider* |
| DMZ | demilitarisierte Zone |
| DWH | *data warehouse* |
| E2E | *end-to end* |
| EAI | *enterprise application integration* |
| EDI | *electronic data interchange* |
| ESB | *Enterprise Service Bus* |
| EU | Europäische Union |
| GB | Gigabyte. 1 GB = 1.024 MB = 1.048.576 kB = 1,073 × 10$^9$ Bytes |
| GSM | *Global System for Mobile Communication*[39] (Mobilfunk 2. Generation) |
| HTTP | *hypertext transfer protocol* |
| HW | *hardware* |
| IaaS | *infrastructure as a service* |
| IEEE | *The Institute of Electrical and Electronics Engineers* |
| IIoT | *industrial IoT* |
| IoT | *Internet oft hings* |

---

[39] In den sehr frühen Anfängen von GSM lautet der Langtext tatsächlich *Group Speciale Mobile*

| Abbreviation | Full text |
| --- | --- |
| IP | *internet protocol* |
| IT | Informationstechnologie |
| KI | künstliche Intelligenz (= AI) |
| km | Kilometer |
| LoRa | *long range* (von LoRaWAN bzw. der LoRa Association |
| LPWAN | *low power WAN* |
| LTE | *Long Term Evolution* (4G Standard) |
| LTE-M | 4G LPWAN Standard |
| M2M | *machine to machine* |
| MB | Megabyte. 1 MB = 1.024 kB = 1.048.576 Bytes |
| MB/s | Megabyte pro Sekunde |
| MES | *Manufacturing Execution System* |
| Mio. | Million. 1 Mio. = 1.000.000 |
| ML | *machine learning* |
| MQTT | *message queuing telemetry transport* |
| Mrd. | Milliarde. 1 Mrd. = 1.000 Mio. = 1.000.000.000 |
| ms | Millisekunde. 1 ms = 1/1.000tel Sekunde |
| NB-IoT | *narrowband IoT* (drahtloses Übertragungsprotokoll in 4G/5G) |
| NR | *new radio* (5G Standard) |
| OPC UA | *Open Platform Communications – Universal Architecture* |
| OS | *operating system* |
| OT | *operations technology* (= AT) |
| P2P | *peer to peer* |
| PaaS | *platform as a service* |
| PLC | *programmable logic array* |
| REST | *representational state transfer* |
| RFID | *radio frequency identification* |
| s | Sekunde |
| SaaS | *software as a service* |
| SDK | *software development kit* |
| SOA | service-orientierte Architektur |
| SPS | Speicherprogrammierbare Steuerung (= PLC) |
| Std. | Stunde |
| SW | *software* |
| TB | Terabyte. 1 TB = 1.204 GB = 1,1 × $10^{12}$ Bytes |
| TCP | *transport control protocol* |
| UMTS | *universal mobile telecommunications system* (Mobilfunk 3. Generation) |
| WAN | *wide area network* |
| WiFi | *wireless fidelity* (WLAN Standard IEEE 802.11) |
| ZB | Zettabyte. 1 ZB = $10^{21}$ Bytes = 1 Mrd. TB |